\documentclass[12pt,a4paper]{article}
\usepackage{color}
\usepackage{epsfig}
\usepackage{amsmath}
\usepackage{amssymb}
\usepackage[english]{babel}
\begin{document}

\begin{center}

\vspace{1cm}

{\bf\large A Note on Efimov Nonlocal and Nonpolynomial Quantum Scalar Field Theory}

\vspace{0.25cm}

{\bf\emph{In Memory of Gariy Vladimirovich Efimov}}

\vspace{0.5cm}

{V.A. Guskov, M.G. Ivanov $^{\dagger}$ and S.L. Ogarkov $^{\ddagger}$}

{Moscow Institute of Physics and Technology (MIPT), Institutskiy Pereulok 9, 141701 Dolgoprudny, Moscow Region, Russia}

{Correspondence: $^{\dagger}$ivanov.mg@mipt.ru, $^{\ddagger}$ogarkovstas@mail.ru}

\end{center}

\begin{abstract}
In frames of the nonlocal and nonpolynomial quantum theory of the one component scalar field in $D$-dimensional spacetime, stated by G.V. Efimov, the expansion of the $\mathcal{S}$-matrix is revisited for different interaction Lagrangians and for some kinds of Gaussian propagators modified by different ultraviolet form factors $F$ which depend on some length parameter $l$. The expansion of the $\mathcal{S}$-matrix is of the form of a grand canonical partition function of some $(D+N)$-dimensional ($N\geq 1$) classical gas with interaction. The toy model of the realistic quantum field theory (QFT) is considered where the $\mathcal{S}$-matrix is calculated in closed form. Then, the functional Schwinger--Dyson and Schr\"{o}dinger equations for the $\mathcal{S}$-matrix in Efimov representation are derived. These equations play a central role in the present paper. The functional Schwinger--Dyson and Schr\"{o}dinger equations in Efimov representation do not involve explicit functional derivatives but involve a shift of the field which is the $\mathcal{S}$-matrix argument. The asymptotic solutions of the Schwinger--Dyson equation are obtained in different limits. Also, the solution is found in one heuristic case allowing us to study qualitatively the behavior of the $\mathcal{S}$-matrix for an arbitrary finite value of its argument. Self-consistency equations, which arise during the process of derivation, are of a great interest. Finally, in the light of the discussion of QFT functional equations, ultraviolet form factors and extra dimensions, the connection with functional (in terms of the Wilson--Polchinski and Wetterich--Morris functional equations) and holographic renormalization groups (in terms of the functional Hamilton--Jacobi equation) is made. In addition the Hamilton--Jacobi equation is formulated in an unconventional way.
\end{abstract}

{\bf Keywords}: Nonlocal and nonpolynomial QFT, grand canonical partition function, functional Schwinger--Dyson and Schr\"{o}dinger equations, functional and holographic RG.

{\bf PACS}: 11.10.-z, 11.55.-m

\newpage

\section{Introduction}\label{ch1}

Historically nonlocal and nonpolynomial quantum field theory (QFT) was developed as an attempt to remove ultraviolet divergences that are connected with the locality of the interaction. After long and often misleading research, the basic concepts, which are necessary to construct a consistent theory of the nonlocal and nonpolynomial interactions of the quantum fields, have been developed. These concepts are still essential even though the theory may be only the first step to construct the more general (hypothetical) and fundamental theory with the fundamental (ultraviolet) length $l$, than quantum field picture of Nature. In this case, nonlocal and nonpolynomial QFT is to play a role of the phenomenological model that fit all the postulates of the QFT where the form factors describe dynamical consequences that we can not calculate from the more general theory on now. If the general theory does not exist and the fundamental length $l$ is only an ultraviolet parameter of the model (as the potential of interaction in the Schr\"{o}dinger equation), it does not affect the construction of nonlocal and nonpolynomial QFT. It is much more interesting to obtain the expressions (for example) for the $\mathcal{S}$-matrix in a form that allows analyzing it as depending on the ultraviolet parameter $l$.

The success of the renormalization theory in local (and, usually, polynomial) QFT and the development of the axiomatic QFT have suppressed the interest in nonlocal QFT. There was a belief that renormalization principle rids of the necessity to formulate a theory without ultraviolet divergences. On the other hand, the interest in the nonpolynomial Lagrangians was still giving a warm hope for the construction of the nonlocal QFT. Eventually, many scientists who worked in this field have understood (see \cite{efimov1970essentially,efimov1970nonlocal,efimov1975proof,petrina1970ksequations,rebenko1972qed,basuev1973conv,basuev1975convYuk} and many others) that the development of one component scalar nonlocal QFT in case of particular nonpolynomial Lagrangians requires {\bf\emph{Euclidean}} metric as an initial point of the theory. It became clear that the ultraviolet form factors depending on the fundamental length $l$ are to be entire analytic functions decreasing in Euclidean space. All these questions are discussed in details in the monographs \cite{efimov1977nonlocal,efimov1985problems}, and in the review \cite{efimov2004blokhintsev}. We note that Efimov's books \cite{efimov1977nonlocal,efimov1985problems} are rich in references to different papers written by the scientists whose research was the development of nonlocal and nonpolynomial QFT. A part of the materials of these books formed the basis for the pedagogical introduction to the classical theory of particles and fields \cite{kosyakov2007intro} because of a deep analysis of the classical side of nonlocal and nonpolynomial field theory in \cite{efimov1977nonlocal,efimov1985problems}. In addition, we also note several original Efimov's papers \cite{efimov1967non,alebastrov1973proof,alebastrov1974proof,efimov1977cmp,efimov1979cmp,efimov2001amplitudes} appeared in the time of rapid development of this problem in QFT.

As it is known now from the methods of the functional, i.e. nonperturbative (or exact), renormalization group, the ultraviolet form factors as functions of the differential operators (in real space, i.e. coordinate space) correspond to the regulators of FRG flow of different generating functionals in QFT and statistical physics \cite{kopbarsch,wipf2012statistical,rosten2012fundamentals,igarashi2009realization}. However, FRG regulators have not been always chosen as entire analytic functions as Efimov suggested. These functions are more appreciable if one wants to obtain FRG flow equations with the best continuity properties. Finally, it should be taken into account that the transition into the Minkowski space cannot be realized through the simple Wick rotation which is possible for local QFT. It is necessary to use other methods of analytical continuation which is a subject of another complicated problem. As a result of the research, in nonlocal and nonpolynomial QFT one can achieve {\bf{\emph{the same level of rigor}}} as in local QFT (for instance, see classical monographs of Bogolyubov \& Shirkov \cite{bogoliubov1980introduction,bogolyubov1983quantum}, and also monographs of Vasiliev \cite{vasil1998methods,vasil2004field} and Zinn-Justin \cite{zinn1989field}).

The main idea of Efimov's papers (see \cite{efimov1970nonlocal} for example) is that {\bf\emph{the interaction action}} $S_{1}$ should be represented in the form that any primary field or composite operator in $S_{1}$ should be an argument of the exponent function. In other words, the construction {\bf\emph{exp in the power of exp}} should appear under the path integral sign. For instance, it is possible if $S_{1}$ has the form of the integral of a primary field {\bf\emph{local}} function with respect to $x$ or of composite operators, i.e. the interaction Lagrangian is local. After that, the generating functional $\mathcal{Z}$ of total Green functions or the $\mathcal{S}$-matrix in Euclidean metric are constructed in terms of the path integral with respect to the primary field. Then, the {\bf\emph{external exp}} (exponential function which contains $S_{1}$ as an argument) should be expanded in Taylor series. As a result, infinite series of Gaussian path integrals are obtained, however each integral can be calculated exactly. As a result of the calculations, final expression for the path integral is found. Of course, calculation of the path integral of {\bf\emph{exp in the power exp}} construction is common in the literature (for example, see discussions of Liouville and sine-Gordon models \cite{polyakov1987gauge,polyakov1977quark,samuel1978grand,o1999duality}). However, it is clear that Efimov representation allows us to calculate the path integral for broad class of theories which will be discussed in details in the next sections.

The expression for the path integral, which is a grand canonical partition function of $D$-dimensional interacting {\bf\emph{quantum}} gas (from the point of view of statistical physics), is obtained in terms of the grand canonical partition function of $(D+N)$-dimensional interacting {\bf\emph{classical}} gas (the value $N$ will be determined further). Then, the interaction potential of the classical gas is defined via quantum field propagator of free theory. This correspondence is also a beautiful example of the duality between (in general, nonconformal) QFT and statistical physics. Unfortunately, the obtained expression for the grand partition function is almost impossible to calculate. For this reason, it is relevant to analyze the functional Schwinger--Dyson and the Schr\"{o}dinger equations in Efimov representation (the Schwinger--Dyson equation for Liouville model is discussed in the interesting paper \cite{dutta2016schwinger}). Also, we note that Efimov devoted substantial attention to functional Schr\"{o}dinger equation. We focus on the Schwinger--Dyson one. All these equations allows us to find several asymptotic solutions for the $\mathcal{S}$-matrix, for example. We pay special attention to the finding of these solutions for $D$-dimensional sine-Gordon theory. This example is already interesting itself since the value of $D$ is not equal to two which is common in literature. In addition, we note that many of the findings obtained for this example remain valid in a more general case.

Discussing the Schwinger--Dyson and the Schr\"{o}dinger equations on the one side and the ultraviolet $l$-depending form factors on the other, we briefly touch upon the FRG equations in the Wilson--Polchinski and the Wetterich--Morris forms (see \cite{kopbarsch,wipf2012statistical,rosten2012fundamentals,igarashi2009realization} for details) and also the discussion of the different coarse-graining procedures of degrees of freedom in a system \cite{rosten2012fundamentals}. This excursus is made for the sake of completeness. We note that in the framework of the nonlocal QFT the value $l^{-1}=\varLambda_{0}$ is considered to be constant and equal to the {\bf\emph{boundary value}} of the FRG flow variable. Also, in light of the appearance of the quantum/classical duality, which in a sense is a holographic picture, initially appeared in completely different quantum field models in papers \cite{maldacena1999large,witten1998anti,gubser1998gauge}, we briefly discuss the functional Hamilton--Jacobi (HJ) equation, which is the main equation in the holographic renormalization group (HRG) method just as the functional Wilson--Polchinski equation underlies one of the possible FRG version. The HRG was developed as a geometrization of the field theoretic renormalization group in \cite{akhmedov1998remark}. The description of the HRG in terms of the HJ equation was presented in papers \cite{de2000holographic,verlinde2000rg}, see \cite{fukuma2003holographic} for more. From the latest (on now) papers in this field, we emphasize the paper \cite{lizana2016holographic} partially devoted to the connection between HRG and FRG methods (also this question is briefly discussed in \cite{akhmedov2003notes,akhmedov2011hints,heemskerk2011holographic}). Our original approach is that we formulate the HJ equation in $D$-dimensional real (coordinate) space $x$. Instead of additional coordinates (extra dimensions), we introduce additional {\bf\emph{holographic}} scalar field (this idea is partially taken from \cite{doplicher2004generalized}). The additional coordinate will appear as the value of the delta-field (in terms of Dirac delta function) configuration of the holographic field. The justification of this approach is that it automatically allows formulating not only the HJ equation but also its expansion into the hierarchy of integro-differential equations for the corresponding (holographic) family of Green functions. A huge advantage of the HJ hierarchy over other hierarchies obtained from the different FRG flow equations is {\bf\emph{decomposition}} of the HJ hierarchy into the independent equations. Consequently, in principle, the HJ hierarchy can be solved in closed form.

To sum up, we emphasize that nonlocal and nonpolynomial QFT formulated by G.V. Efimov allows us to analyze the strong coupling regime in different quantum field theories, and any strong coupling method in modern QFT is still as good as gold. Also, we should note that from the mathematical point of view the structure of nonlocal and nonpolynomial QFT is close to the objects studied in statistical physics. The appearance of new results at the crossover between two these fundamental sciences is just a matter of time. We sincerely believe that methods of the nonlocal and nonpolynomial QFT will occupy a worthy place in description of Nature.

\section{$\mathcal{S}$-matrix in Efimov representation}\label{ch2}

\subsection{Zoology of the nonpolynomial Lagrangians in nonlocal QFT and nested exponentials}

The first principle of quantum field theory and statistical physics is the formulation of the (nonnormalized) generating functional $\mathcal{Z}$ of the family of total (including disconnected diagrams) Green functions in terms of the path integral over the primary field $\varphi$ (in the Euclidean metric, like in the monographs \cite{kopbarsch}, \cite{vasil1998methods,vasil2004field} or \cite{zinn1989field}):
    \begin{eqnarray}\label{ch2-eq1}
        \mathcal{Z}\left[g,j\right]=\int\mathcal{D}\left[\varphi\right]
        e^{-S\left[g,\varphi\right]+\left(j\mid\varphi\right)}.
    \end{eqnarray}
The action $S$ that depends not only on primary field but also on some function $g$, the meaning of which will be clear in a moment, is usually formed by  the free theory $S_{0}$ and the interaction $S_{1}$ actions:
    \begin{eqnarray}\label{ch2-eq2}
        S\left[g,\varphi\right]=S_{0}\left[\varphi\right]+S_{1}\left[g,\varphi\right].
    \end{eqnarray}
In nonlocal QFT the expression for the free theory action $S_{0}$ reads as ($m$ is an infrared mass parameter, $l$ is an ultraviolet length parameter, $F$ is an ultraviolet form factor ensuring the convergence of integrals at large momentum):
    \begin{eqnarray}\label{ch2-eq3} 
        S_{0}\left[\varphi\right]=\frac{1}{2}\int d^{D}x\,\varphi\left(x\right)
        L\left(\partial\right)\varphi\left(x\right),\quad
        L^{-1}\left(\partial\right)=
        \frac{F\left(-l^{2}\partial^{2}\right)}{-\partial^{2}+m^{2}}.
    \end{eqnarray}
It is essential to mention here that the form factor $F$ is not designed to make any integrals we want finite but to make the exact expression for generating functional $\mathcal{Z}$ finite. Also because of the form factor $F$ we are not limited to choose the spacetime dimension $D$ less or equal to $4$. That is why the choice of $D=4$ in this paper has only an illustrative character.

The scalar product of the source field $j$ and the primary field $\varphi$ has a standard form (in the second expression we may see the typical form of $j$ if we are interested in Green functions of exponential operators -- that is a typical choice for the Liouville model, for instance; $\alpha_{a}$ and $z_{a}$ are constant parameters):
    \begin{eqnarray}\label{ch2-eq4}
        \left(j\mid\varphi\right)=\int d^{D}x\,j\left(x\right)
        \varphi\left(x\right),\quad
        j\left(x\right)=\sum\limits_{a=1}^{N}
        \alpha_{a}\,\delta^{\left(D\right)}
        \left(x-z_{a}\right).
    \end{eqnarray}
Finally, let us define the interaction action $S_{1}$. In the general case it may be defined by the field and its derivatives in many different ways. In this paper we assume that under the integral sign there is the local function of field $\varphi$. In other words, the interaction Lagrangian is considered to be local ($g$ is the function that describes a smooth ``turning-on'' and ``turning-off'' of the corresponding interaction):
    \begin{eqnarray}\label{ch2-eq5}
        S_{1}\left[g,\varphi\right]=\int d^{D}x\,g\left(x\right)U\left[\varphi\left(x\right)\right],\quad
        U\left[\varphi\left(x\right)\right]=\int\limits_{-\infty}^{\infty}\frac{d\lambda}{2\pi}\,
    \tilde{U}\left(\lambda\right)e^{i\lambda\varphi\left(x\right)}.
    \end{eqnarray}
In the second expression (\ref{ch2-eq5}) it is assumed that the interaction Lagrangian may be written in a form of a standard Fourier integral over an infinite real axis (where $\tilde{U}$ denotes the Fourier transform of $U$).

In the framework of the theory developing in our paper this choice is natural but not the only one. For instance, for the polynomial interaction Lagrangian $\varphi^{n}$ it is better to use the Fourier transform in a complex plane of the variable $\lambda$ over the unit circle. In this case, a rapidly divergent perturbation theory (PT) series arises, but this problem can be solved, for example, in the spirit of papers \cite{basuev1974summPT,BelokurovI,BelokurovII}, and especially \cite{korsun1992VPT}. In addition, it is possible to consider singular interactions and here we need to use Fourier transform in the vicinity of regular point. The following expressions demonstrate what have been discussed (see monographs \cite{efimov1977nonlocal} and \cite{efimov1985problems}):
    \begin{eqnarray}\label{ch2-eq6}
    \varphi^{n}\left(x\right)=
    \frac{\left(-i\right)^{n}n!}{2\pi i}\oint\limits_{|\lambda|=1}\frac{d\lambda}
    {\lambda^{n+1}}\,
    e^{i\lambda\varphi\left(x\right)},\quad
    \frac{1}{\sqrt{1-\varphi^{2}\left(x\right)}}=
    \frac{1}{2}
    \int\limits_{-\infty}^{\infty}d\lambda\,
    J_{0}\left(\lambda\right)
    e^{i\lambda\varphi\left(x\right)}.
    \end{eqnarray}
Approaches presented in this paper allow a direct generalization in the case when in the expression for $S_{1}$ under the integral sign there is the local Lagrangian depending not only on the field $\varphi$ but also on some composite operators. This generalization only increases the number of Fourier variables which are necessary to put the field $\varphi$ and all the composite operators in the exponent.
    \begin{eqnarray}\label{ch2-eq7}
        U\left[\varphi\left(x\right)\right]\rightarrow
    U\left[\varphi\left(x\right),O_{1}\left(\partial\right)\varphi\left(x\right),\ldots,
    O_{N}\left(\partial\right)\varphi\left(x\right)\right].
    \end{eqnarray}
The increase in the number of primary fields is also a related generalization. For instance, one can consider $O(N)$-symmetrical model. All the fields $\varphi_{1},\ldots,\varphi_{N}$ in the interaction should be dispatch to the exponent:
    \begin{eqnarray}\label{ch2-eq8}
        U\left[\varphi\left(x\right)\right]\rightarrow
        U\left[\varphi_{1}\left(x\right),\ldots,\varphi_{N}\left(x\right)\right]\rightarrow
        \tilde{U}\left(\lambda_{1},\ldots,\lambda_{N}\right).
    \end{eqnarray}
This correspondence between the amount of different ``building blocks'' in the action and the amount of Fourier variables will get an important holographic interpretation in the future. 

At the end of the excursus to the zoology of permitted interactions let us notice one exotic but interesting case. The theory presented in this paper allows us to consider field theories in which the interaction action occurs not in the exponent but as an argument of some arbitrary function $f$ with good convergence properties. \emph{It is important that such a function can be represented by a Taylor series converging to it in some vicinity of zero.} One can consider entire function as an example. The following equalities demonstrate this:
    \begin{eqnarray}\label{ch2-eq9}
        e^{-S_{1}\left[g,\varphi\right]}\rightarrow f\left\{S_{1}\left[g,\varphi\right]\right\}=
        \sum\limits_{n=0}^{\infty}\frac{f_{n}}{n!}\left\{S_{1}
        \left[g,\varphi\right]\right\}^{n}.
    \end{eqnarray}
In conclusion of this brief introduction let us also notice that we will often illustrate final results in terms of the sine-Gordon field theory (in this theory the parameter $\alpha\in\mathbb{R}$ is assumed to be fixed):
    \begin{eqnarray}\label{ch2-eq10}
    S_{1}\left[g,\varphi\right]=2\int d^{D}x\,g\left(x\right)
    \cos\left[\alpha\varphi\left(x\right)\right]=
    \int d^{D}x\,g\left(x\right)
    \left\{e^{i\alpha\varphi\left(x\right)}+
    e^{-i\alpha\varphi\left(x\right)}\right\}.
    \end{eqnarray}
Such approach does not influence the general character of the obtained results. It is enough to note that in the expression (\ref{ch2-eq10}) the dimension of spacetime $D$ is arbitrary although only $D=2$ case is often found in the literature (see \cite{polyakov1987gauge} or \cite{samuel1978grand}).

\subsection{Structure of the propagator in nonlocal QFT and its duality to the Lagrangian of the nonlocal interaction}

In this subsection we consider in detail the expression (\ref{ch2-eq3}) for propagator of the free theory $S_{0}$ in nonlocal QFT. Such expression is typical, for example, for the Gaussian $\varLambda$-deformed (by a regulator $R_{\varLambda}$, where $\varLambda$ is the running momentum scale) propagator which is used in functional renormalization group (FRG) method. A brilliant description of FRG methods can be found in the monographs \cite{kopbarsch,wipf2012statistical} and the reviews \cite{rosten2012fundamentals,igarashi2009realization}. The value $\varLambda$ launches the FRG flow formulating different evolutionary equations in functional derivatives, for instance, the Wilson--Polchinski or the Wetterich--Morris equations. In terms of nonlocal QFT the value $l^{-1}=\varLambda_{0}$, where $\varLambda_{0}$ is the ultraviolet momentum scale, is assumed to be fixed and equal to boundary value of the flow variable. Thus, the modification of the Gaussian propagator by the ultraviolet form factor, depending on $l$, is a typical construction in quantum field theory. Let us consider this construction in the momentum representation:
    \begin{eqnarray}\label{ch2-eq11}
        G\left(x\right)=\int_{k}\frac{F\left(w\right)e^{ikx}}{k^{2}+m^{2}}
        ,\quad\int_{k}=\int\frac{d^{D}k}{\left(2\pi\right)^{D}},\quad w=l^{2}k^{2}.
    \end{eqnarray}
In the case of the four-dimensional spacetime ($D=4$), which is of special interest for QFT, the expression (\ref{ch2-eq11}) can be rewritten in the following compact form:
    \begin{eqnarray}\label{ch2-eq12}
        G\left(x\right)=\int\limits_{0}^{\infty}\frac{du}{\left(2\pi\right)^{2}}\,\frac{uF\left(w\right)}
        {u+m^{2}}\frac{J_{1}\left(\sqrt{ux^{2}}\right)}{2\sqrt{ux^{2}}}.
    \end{eqnarray}
For the sake of completeness we introduce few examples of the form factor $F$. The general analysis of the functions $F$ is made in details in Efimov's papers \cite{efimov1970essentially,efimov1970nonlocal,efimov1975proof} and in his books \cite{efimov1977nonlocal,efimov1985problems}. Also it is possible to use the results from FRG methods which are appropriate due to the coincidence of the ultraviolet form factor $F$ and the FRG regulator $R_{\varLambda}$. The examples of the form factors are given by (we can improve the convergence at infinity if it is necessary):
    \begin{eqnarray}\label{ch2-eq13}
        F_{1}\left(w\right)=e^{-w},\quad
        F_{2}\left(w\right)=\left(\frac{\sin\sqrt{w}}
        {\sqrt{w}}\right)^{2},\quad
        F_{3}\left(w\right)=\frac{9}{w^{2}}\left(\frac{\sin\sqrt{w}}
        {\sqrt{w}}-\cos\sqrt{w}\right)^{2}.
    \end{eqnarray}
To illustrate the general results we will use the first example in this paper. 
Now we pay attention to the following important feature: the duality of the propagator structure in nonlocal QFT to the Lagrangian of the nonlocal interaction. All the results which are appropriate for the modified Gaussian propagator can be derived from an another approach: we may use the standard propagator for the local QFT but modify the interaction Lagrangian $U$ in (\ref{ch2-eq5}). This modification is given by:
\begin{eqnarray}\label{ch2-eq14}
    F\left(w\right)=\left[K\left(w\right)\right]^{2}\rightarrow
    U\left[K\left(l^{2}\partial^{2}\right)\varphi\left(x\right)\right].
\end{eqnarray}
In other words, we need to consider a corresponding composite operator $K\varphi$ as an argument of the Lagrangian $U$. To prove this statement we need to make a corresponding substitution of the primary field under the path integration sign in the expression (\ref{ch2-eq1}).

In conclusion of this subsection we note several important inequalities for further use. Following the fundamental Efimov's paper \cite{efimov1970nonlocal} it is easy to show that for real values of $\lambda_{a}$ and $\lambda_{b}$ the modified by the form factor propagator of free theory forms a positive definite quadratic form:
    \begin{eqnarray}\label{ch2-eq15}
        \sum\limits_{a,b=1}^{n}\lambda_{a}\lambda_{b}G\left(x_{a}-x_{b}\right)=
        \int_{k}\frac{F\left(w\right)}{k^{2}+m^{2}}\bigg|\sum\limits_{a=1}^{n}
        \lambda_{a}e^{ikx_{a}}\bigg|^{2}\geq0.
    \end{eqnarray}
Moreover, formed by the propagator $G$ the quadratic form is bounded in accordance with Cauchy--Schwarz--Bunyakovsky inequality. Thus, we have one more inequality:
    \begin{eqnarray}\label{ch2-eq16}
        \sum\limits_{a,b=1}^{n}\lambda_{a}\lambda_{b}G\left(x_{a}-x_{b}\right)\leq
        nG\left(0\right)\sum\limits_{a=1}^{n}\lambda_{a}^{2}.
    \end{eqnarray}
The inequalities (\ref{ch2-eq15})--(\ref{ch2-eq16}) will play an important role in the proof of the fact that generating functional  $\mathcal{Z}$ is finite in nonlocal and nonpolynomial QFT formulated in this paper following Efimov.

\subsection{$\mathcal{S}$-matrix in Efimov representation}

In this subsection we consider the most important functional in QFT which is the $\mathcal{S}$-matrix of the theory. Like the generating functional $\mathcal{Z}$ of total Green functions, the $\mathcal{S}$-matrix (being the functional of some field $\varphi$) is also a generating functional of some family of Green functions. As is known from standard QFT courses the logarithm of the $\mathcal{S}$-matrix $\mathcal{G}$ is the generating functional of the amputated (without Gaussian propagators with external coordinates $x$ or momenta $k$) connected Green functions. The following expressions illustrate this (the primary field in the functional integral we denoted as $\psi$ in order to think of field $\varphi$ as an argument of the $\mathcal{S}$-matrix, like in monographs \cite{kopbarsch}, \cite{vasil1998methods,vasil2004field} or \cite{zinn1989field}):
    \begin{eqnarray}\label{ch2-eq17}
        \mathcal{S}\left[g,\varphi\right]=e^{\mathcal{G}\left[g,\varphi\right]}=\frac{1}{\mathcal{Z}_{0}}
        \int\mathcal{D}\left[\psi\right]e^{-S_{0}\left[\psi\right]-
        S_{1}\left[g,\psi+\varphi\right]}.
    \end{eqnarray}
The partition function of the free theory $\mathcal{Z}_{0}$ in the expression (\ref{ch2-eq17}) is a standard normalization of the $\mathcal{S}$-matrix. Finally, in the expression (\ref{ch2-eq17}) it is taken into account that the $\mathcal{S}$-matrix is a functional of the coupling constant $g$, where $g$ is the function which describes smooth ``turning-on'' and ``turning-off'' of the interaction.

It is well-known (see monographs \cite{bogoliubov1980introduction} and \cite{bogolyubov1983quantum}) that the $\mathcal{S}$-matrix can be represented as a result of the action of an (exponential) series of functional derivatives with respect to the field $\varphi$ on the exponent to the power of interaction action $S_{1}$:
    \begin{eqnarray}\label{ch2-eq18}
        \mathcal{S}\left[g,\varphi\right]=
        e^{\frac{1}{2}\left(\frac{\delta}{\delta\varphi}\mid
        \mathbf{G}\frac{\delta}{\delta\varphi}\right)}
        e^{-S_{1}\left[g,\varphi\right]}.
    \end{eqnarray}
This expression is a starting point for the construction of the perturbation theory in the framework of local QFT. Moreover, precisely the differential operator in (\ref{ch2-eq18}) creates the main problem of the PT series which is the asymptotic property.

Being the functional of the field $\varphi$, the $\mathcal{S}$-matrix may be expanded in a functional Taylor series with respect to the field configurations $\varphi$:
    \begin{eqnarray}\label{ch2-eq19}
        \mathcal{S}\left[g,\varphi\right]=
        \sum\limits_{n=0}^{\infty}
        \frac{1}{n!}
        \int d^{D}x_{1}\ldots\int d^{D}x_{n}\,\mathcal{S}^{(n)}\left[g\right]
        \left(x_{1},\ldots,x_{n}\right)
        \varphi\left(x_{1}\right)\ldots
        \varphi\left(x_{n}\right).
    \end{eqnarray}
At the same time, as the functional of the coupling constant $g$ the $\mathcal{S}$-matrix may be expanded into a functional Taylor series over $g$:
    \begin{eqnarray}\label{ch2-eq20}
        \mathcal{S}\left[g,\varphi\right]=
        \sum\limits_{n=0}^{\infty}
        \frac{\left(-1\right)^{n}}
        {n!}\!\!\int d^{D}x_{1}\ldots\int d^{D}x_{n}\,\mathcal{S}_{n}\left[\varphi\right]
        \left(x_{1},\ldots,x_{n}\right)
        g\left(x_{1}\right)\ldots g\left(x_{n}\right).
    \end{eqnarray}
The last expansion is under detailed discussions in classical monographs \cite{bogoliubov1980introduction,bogolyubov1983quantum}.

In Efimov's paper \cite{efimov1970nonlocal} it is shown that the correlation functions $\mathcal{S}_{n}$ in (\ref{ch2-eq20}) may be calculated in a closed form. The idea is very simple and elegant: we need to send the field in the expression for the interaction action to the exponent, i.e. we need to obtain nested exponentials, and then expand the external exponent (with $S_{1}$ as an argument) into the Taylor series. As a result we obtain an infinite series of Gaussian functional integrals but each of them may be calculated by the well-known formulas. As a result we obtain the following result:
    \begin{eqnarray}\label{ch2-eq21}
        \mathcal{S}_{n}\left[\varphi\right]\left(\left\{x\right\}\right)=\int\limits_{-\infty}^{\infty}\frac{d\lambda_{1}}{2\pi}\ldots\int\limits_{-\infty}^{\infty}\frac{d\lambda_{n}}{2\pi}\,
        \tilde{U}\left(\lambda_{1}\right)\ldots\tilde{U}\left(\lambda_{n}\right)\times\nonumber\\ \times\exp\left\{-\frac{1}{2}\sum\limits_{a,b=1}^{n}\lambda_{a}\lambda_{b}G\left(x_{a},x_{b}\right)+i\sum\limits_{a=1}^{n}\lambda_{a}\varphi\left(x_{a}\right)\right\}\,.
    \end{eqnarray}
Thus, the expressions (\ref{ch2-eq20})--(\ref{ch2-eq21}) give us a final answer for the $\mathcal{S}$-matrix in terms of the corresponding series. 

Here we should stop for a moment and answer the following question: how far should we consider the obtained expressions as the answers? To answer this question we need to remember, what is the origin of the functional integrals like (\ref{ch2-eq1}) and (\ref{ch2-eq17}) (for example) in statistical physics. The functional integral provides a way to write down the grand canonical partition function of the $D$-dimensional quantum gas with interaction. Moreover, the expressions (\ref{ch2-eq20})--(\ref{ch2-eq21}) are the grand canonical partition functions of the $D+1$-dimensional classical gas with an interaction given by the quantum field propagator $G$. If an additional $N-1$ composite operator was in the interaction action $S_{1}$, the dimension of the extended space for classical gas would be $D+N$. So, we have a specific holographic picture: the $D$-dimensional quantum system may be presented in terms of the effective $(D+N)$-dimensional classical one.

In this context it is worth discussing the physical meaning  of the additional coordinate $\lambda$ (in a more general case, a set of additional coordinates $\lambda_{1},\ldots,\lambda_{N}$). In QFT and statistical physics the field $\varphi$ (as so as composite operators) plays the role of the functional argument. That is why $\varphi$ defines the domain of the theory. The argument of the field $\varphi$, either $x$ or $k$, plays the role of the index. It turns out that in the case when the interaction action $S_{1}$ has the form like (\ref{ch2-eq5}) the domain of the theory is just a unification of $x$ and $\lambda$. The meaning of the latter is the range of the field $\varphi$ (direct connection between $\lambda$ and $\varphi$ is performed by the Fourier integral). So, in Efimov theory the field $\varphi$ has a meaning of an extra dimension much more then a function of $x$. In some sense this reproduces a situation in quantum mechanics (QM) where a trajectory $q(t)$ used in formulation of QM in terms of a functional integral becomes a variable in the Schr\"{o}dinger equation.

Now let us explore the convergence properties of the expansion (\ref{ch2-eq20})--(\ref{ch2-eq21}). In the paper \cite{efimov1970nonlocal} it is shown that for this expansion there is a majorizing series (majorant) which does not depend on $\varphi$:
    \begin{eqnarray}\label{ch2-eq22}
        \mathcal{S}_{M}\left[g\right]=\exp\Bigg\{\int d^{D}x\,g\left(x\right)
        \int\limits_{-\infty}^{\infty}\frac{d\lambda}{2\pi}\,\big|\tilde{U}
        \left(\lambda\right)\big|\Bigg\}.
    \end{eqnarray}
Let us note here that \emph{the expression (\ref{ch2-eq22}) is the result of summation of the majorizing series, and not its term.} It means that the expansion (\ref{ch2-eq20})--(\ref{ch2-eq21}) converges absolutely and this is the main result of the paper \cite{efimov1970nonlocal}. Despite the fact that the derivation is correct for QFT in the Euclidean formulation (therefore it is correct and sufficient for statistical physics) the result should be considered as a starting point for the construction of QFT in Minkowski space. Although, the analytic continuation of the obtained results in Minkowski space should be an independent subject for study. As it was stated by Efimov, this continuation can not be carried out by the Wick rotation of the integration path in the momentum space. Let us also note that the series of this type can be obtained by evaluating the Green functions of the theory.

At the end of this subsection let us give the $\mathcal{S}$-matrix of the sine-Gordon field theory as an example for (\ref{ch2-eq20})--(\ref{ch2-eq21}). The simplest strategy for the derivation of the $\mathcal{S}$-matrix for such a theory is as follows: we start with action (\ref{ch2-eq10}) in terms of two exponents. Next, we write the following expression for the exponent containing the interaction action $S_{1}$ of the system with a minus sign (we introduce two sets of vectors $x$ and $y$ to distinguish between two exponents):
\begin{eqnarray}\label{Reviewer-1}
    e^{-S_{1}\left[g,\varphi\right]}=
    \sum\limits_{n,m=0}^{\infty}
    \frac{\left(-1\right)^{n+m}}{n!m!}
    \int d^{D}x_{1}
    \ldots\int d^{D}x_{n}\int d^{D}y_{1}\ldots\int d^{D}y_{m}\times\nonumber\\
    \times g\left(x_{1}\right)\ldots g\left(x_{n}\right)
    g\left(y_{1}\right)\ldots g\left(y_{m}\right)
    e^{i\alpha\left\{\varphi\left(x_{1}\right)+\ldots
    +\varphi\left(x_{n}\right)
    -\varphi\left(y_{1}\right)-\ldots
    -\varphi\left(y_{m}\right)\right\}}.
\end{eqnarray}
Using the expression (\ref{Reviewer-1}), we can calculate the $\mathcal{S}$-matrix in terms of the functional integral over the field $\psi$ (\ref{ch2-eq17}). Thus, we arrive at the following expression:
    \begin{eqnarray}\label{ch2-eq23}
        \mathcal{S}\left[g,\varphi\right]=\sum\limits_{n,m=0}^{\infty}
        \frac{\left(-1\right)^{n+m}}{n!m!}\int d^{D}x_{1}
        \ldots\int d^{D}x_{n}\int d^{D}y_{1}\ldots\int d^{D}y_{m}\times\nonumber\\
        \times
        g\left(x_{1}\right)\ldots g\left(x_{n}\right)
        g\left(y_{1}\right)\ldots g\left(y_{m}\right)\mathcal{S}_{n,m}\left[\varphi\right]
        \left(\left\{x\right\},\left\{y\right\}\right).
    \end{eqnarray}
The expression for the correlation functions $\mathcal{S}_{n,m}$ reads:
    \begin{eqnarray}\label{ch2-eq24}
        \mathcal{S}_{n,m}\left[\varphi\right]
        \left(\left\{x\right\},\left\{y\right\}\right)=
        \exp\left\{\alpha^{2}
        \sum\limits_{a=1}^{n}\sum\limits_{b=1}^{m}
        G\left(x_{a},y_{b}\right)\right\}
        \times\nonumber\\
        \times\exp\left\{-\frac{\alpha^{2}}{2}\!
        \sum\limits_{a,a'=1}^{n}\!G
        \left(x_{a},x_{a'}\right)-
        \frac{\alpha^{2}}{2}\!
        \sum\limits_{b,b'=1}^{m}\!
        G\left(y_{b},y_{b'}\right)\right\}
        \times\nonumber\\
        \times\exp\left\{i\alpha
        \sum\limits_{a=1}^{n}\varphi
        \left(x_{a}\right)-i\alpha
        \sum\limits_{b=1}^{m}\varphi
        \left(y_{b}\right)\right\}.
    \end{eqnarray}
In the case of $D=2$, the $\mathcal{S}$-matrix (\ref{ch2-eq23})--(\ref{ch2-eq24}) was the subject of an intensive investigation in the literature (see \cite{polyakov1987gauge}, \cite{samuel1978grand}, and \cite{o1999duality}).

\subsection{$\mathcal{S}$-matrix in case of separable propagator: zero-dimensional QFT} 

In this subsection we provide a simple example (zero-dimensional QFT) which illustrates the theory above. In the framework of this example we can obtain the exact expression for the $\mathcal{S}$-matrix. Let us consider the case when the propagator $G$ is separable ($\sigma_{s}$ are the functions of a separable basis). 
    \begin{eqnarray}\label{ch2-eq25}
        G\left(x_{a},x_{b}\right)=\sum\limits_{s=1}^{N}
        \sigma_{s}\left(x_{a}\right)\sigma_{s}\left(x_{b}\right).
    \end{eqnarray}
In the expression (\ref{ch2-eq25}) we use Hubbard--Stratonovich transformation for ordinary integrals ($t_{s}$ are auxiliary parameters of Hubbard--Stratonovich):
    \begin{eqnarray}\label{ch2-eq26}
        \exp{\left\{-\frac{1}{2}\sum\limits_{a,b=1}^{n}\lambda_{a}\lambda_{b}G\left(x_{a},x_{b}\right)\right\}}= \prod\limits_{s=1}^{N}\,\int\limits_{-\infty}^{\infty}\frac{dt_{s}}{\sqrt{2\pi}}\exp{\left\{-\frac{1}{2}t_{s}^{2}-it_{s}\sum\limits_{a=1}^{n}\lambda_{a}\sigma_{s}\left(x_{a}\right)\right\}}.
    \end{eqnarray}
Having all the calculations performed the expression for the $\mathcal{S}$-matrix is given by:
    \begin{eqnarray}\label{ch2-eq27}
        \mathcal{S}\left[g,\varphi\right]= \int\limits_{-\infty}^{\infty}\frac{dt_{1}}{\sqrt{2\pi}}\ldots\int\limits_{-\infty}^{\infty}\frac{dt_{N}}{\sqrt{2\pi}}
        \exp{\left\{-\frac{1}{2}\sum\limits_{s=1}^{N}t_{s}^{2}-\int d^{D}x\,g\left(x\right)U\left[\varPhi\left(x,\left\{t\right\}\right)\right]\right\}}.
    \end{eqnarray}
The argument $\varPhi$ of the interaction Lagrangian $U$ in the expression (\ref{ch2-eq27}) is given by the following composite value:
    \begin{eqnarray}\label{ch2-eq28}
        \varPhi\left(x,t_{1},\left\{t\right\}\right)=\varphi\left(x\right)-
        \sum\limits_{s=1}^{N}t_{s}\sigma_{s}\left(x\right).
    \end{eqnarray}
For this zero-dimensional QFT (\ref{ch2-eq27})--(\ref{ch2-eq28}) it is also possible to consider various field configurations or correlation functions. The given example is useful because it allows to develop intuition in how the answers should look like in the case of the $D$-dimensional QFT.

Let us make an important comment here: the Hubbard--Stratonovich transformation may be done in the general expression (\ref{ch2-eq20})--(\ref{ch2-eq21}) for the $\mathcal{S}$-matrix. In the framework of the local but nonpolynomial QFT this approach is developed in the fundamental papers of Efim Samoylovich Fradkin \cite{fradkin1963application,fradkin1966application} (see also monograph \cite{fradkin2007selected}), and the corresponding representation of the $\mathcal{S}$-matrix is called Efimov--Fradkin representation (according to the independent research in papers \cite{efimov1963construction,efimov1965nonlinear}, see also \cite{efimov1977nonlocal} as well as papers of other authors associated with the mentioned \cite{volkov1968cmp,volkov1969cmp,zumino1969,salam1969,salam1970,fivel1970,graffi1972,horvath1972,wataghin1973,sudarshan1973}). At the same time, Efimov representation is more convenient, for instance, for the derivation and the analysis of the functional Schwinger--Dyson and Schr\"{o}dinger equations. That is why Efimov--Fradkin representation will not be discussed further. Nevertheless, this representation may be a subject for another publication because it is important to revisit such quantum field constructions over time.

\subsection{Toy model for realistic QFT}

In the final subsection of this section we take a look at the toy model which connects the realistic $D$-dimensional QFT with the propagator (\ref{ch2-eq11}) and the zero-dimensional QFT with a separable propagator (\ref{ch2-eq25}) which we have discussed earlier. For this purpose let us rewrite the quadratic form with the propagator $G$ from (\ref{ch2-eq11}):
    \begin{eqnarray}\label{ch2-eq29}
        \sum\limits_{a,b=1}^{n}\lambda_{a}\lambda_{b}G\left(x_{a}-x_{b}\right)=\int_{k}\frac{F\left(k\right)}{k^{2}+m^{2}}
        \left[C^{2}\left(k\right)+S^{2}\left(k\right)\right].
    \end{eqnarray}
In the expression (\ref{ch2-eq29}) we introduced the following notations:
    \begin{eqnarray}\label{ch2-eq29sub}
        C\left(k\right)=\sum\limits_{a=1}^{n}\lambda_{a}
        \cos{\left(kx_{a}\right)},\quad
        S\left(k\right)=\sum\limits_{a=1}^{n}\lambda_{a}
        \sin{\left(kx_{a}\right)}.
    \end{eqnarray}
Let us choose the non-Efimov form factor as follows ($f_{s}$ and $k_{s}$ are constant parameters of the model):
    \begin{eqnarray}\label{ch2-eq30}
        F\left(k\right)=\frac{1}{2}\sum\limits_{s=1}^{N}f_{s}\left(2\pi\right)^{D}
        \left[\delta^{\left(D\right)}\left(k-k_{s}\right)+\delta^{\left(D\right    )}
        \left(k+k_{s}\right)\right].
    \end{eqnarray}
In this case we obtain a field theory with a separable propagator (\ref{ch2-eq25}) for which the functions of the separable basis $\sigma_{s}$ are given by:
    \begin{eqnarray}\label{ch2-eq31}
        \sigma^{\left(1\right)}_{s}\left(x\right)=\sqrt{f'_{s}}
        \cos{\left(k_{s}x\right)},\quad\sigma^{\left(2\right)}_{s}\left(x\right)=\sqrt{f'_{s}}\sin{\left(k_{s}x\right)},\quad
        f'_{s}=\frac{f_{s}}{k_{s}^{2}+m^{2}}.
    \end{eqnarray}
Thus, the non-Efimov form factor (\ref{ch2-eq30}) connects the realistic and the toy quantum field theories. 

Also it should be noted that the toy model, considered above, has more in common with the realistic QFT, than it might be seem from the first sight. Indeed, let us consider the expression (\ref{ch2-eq29}). We use the Jensen inequality for convex functions, in this case, for squares of functions $C$ and $S$ defined in (\ref{ch2-eq29sub}). As a result, we arrive to the fact that the quadratic form in (\ref{ch2-eq29}) is bounded from below by a form constructed with the help of a separable propagator, which exactly coincides with $G$:
    \begin{eqnarray}\label{ch2-new1}
        \sum\limits_{a,b=1}^{n}\lambda_{a}\lambda_{b}
        G\left(x_{a}-x_{b}\right)\geq \frac{1}
        {G\left(0\right)}\left\{\sum\limits_{a=1}^{n}\lambda_{a}
        G\left(x_{a}\right)\right\}^{2}.
    \end{eqnarray}
Then, we again use the Hubbard--Stratonovich transformation in (\ref{ch2-new1}) to decouple the dependence on $a$ and $b$ in the exponent:
    \begin{eqnarray}\label{ch2-new2}
       \exp{\left\{-\frac{1}{2}\sum\limits_{a,b=1}^{n}
       \lambda_{a}\lambda_{b}G\left(x_{a},x_{b}\right)\right\}}\leq \int\limits_{-\infty}^{\infty}\!\!\frac{dt}
        {\sqrt{2\pi}}\exp{\left\{-\frac{1}{2}t^{2}-\frac{it}
        {\sqrt{G\left(0\right)}}\sum\limits_{a=1}^{n}\lambda_{a}
        G\left(x_{a}\right)\right\}}.
    \end{eqnarray}
Using expression (\ref{ch2-new2}), it is easy to obtain the following majorizing series (majorant) for the $\mathcal{S}$-matrix of the theory (and it still does not depend on the field $\varphi$):
    \begin{eqnarray}\label{ch2-new3}
        \mathcal{S}_{M}\left[g\right]=\int\limits_{-\infty}^{\infty}
        \frac{dt}{\sqrt{2\pi}}\exp{\left\{-\frac{1}{2}t^{2}+\int d^{D}x\,g\left(x\right)U_{M}\left(x\right)\right\}}.
    \end{eqnarray}
Function $U_{M}$ in expression (\ref{ch2-new3}) is defined as follows:
    \begin{eqnarray}\label{ch2-new4}
       U_{M}\left(x\right)=\int\limits_{-\infty}^{\infty}
        \frac{d\lambda}{2\pi}\,\big|\tilde{U}
        \left(\lambda\right)\big|
        \exp{\left\{\frac{it\lambda G\left(x\right)}
        {\sqrt{G\left(0\right)}}\right\}}.
    \end{eqnarray}
Thus, we can conclude that the separable approximation is a good starting point for calculating of the $\mathcal{S}$-matrix corresponding to the realistic QFT. At the same time, the correlation effects will only lower the value of the $\mathcal{S}$-matrix in the separable approximation up to the realistic ones.

It is possible to find the answer for the $\mathcal{S}$-matrix of nonlocal and nonpolynomial QFT not in terms of the series (\ref{ch2-eq20})--(\ref{ch2-eq21}) but in terms of the asymptotic expressions obtained from corresponding asymptotics of functional equations. This equations may be, for instance, the Schwinger--Dyson equation or the functional Schr\"{o}dinger equation to the discussion of which we proceed in the next section.

\section{Schwinger--Dyson and functional Schr\"{o}dinger equations}\label{ch3}

\subsection{Derivation of the equations in Efimov representation}

The term Schwinger--Dyson (SD) equations is generally used for any relations expressing the equality to zero of the functional integral of the first functional derivative of a given functional (we consider zero boundary conditions at infinity). The brilliant presentation of Schwinger--Dyson equations can be found in monographs \cite{kopbarsch} and \cite{vasil1998methods,vasil2004field}. For instance, there is a relation for the functional which is the integrand in (\ref{ch2-eq1}):
    \begin{eqnarray}\label{ch3-eq1}
        \int\mathcal{D}\left[\varphi\right]\frac{\delta e^{-S\left[g,\varphi\right]+\left(j\mid\varphi\right)}}
        {\delta\varphi\left(x\right)}=\int\mathcal{D}
        \left[\varphi\right]e^{-S\left[g,\varphi\right]+
        \left(j\mid\varphi\right)}\left\{-\frac{\delta S\left[g,\varphi\right]}{\delta\varphi\left(x\right)}+
        j\left(x\right)\right\}=0.
    \end{eqnarray}
First of all, the primary field $\varphi$ functional derivative of the full action $S$ should be calculated. In contrast to a standard derivation of the SD equation, for interaction action $S_{1}$ we use Efimov representation. Then, we use the source trick (with respect to $j$). The primary field $\varphi$ functional derivative of the full action $S$ becomes an operator in terms of the source $j$ functional derivatives. Moreover, in contrast to standard types of the SD equations, the obtained operator is the functional translation operator. The next equalities demonstrate this derivation:
    \begin{eqnarray}\label{ch3-eq2}
        \frac{\delta S\left[g,\varphi\right]}{\delta\varphi\left(x\right)}=
        L\left(\partial\right)\varphi\left(x\right)+g\left(x\right)
        \int_{\lambda}\tilde{U}\left(\lambda\right)i\lambda\,
        e^{i\lambda\varphi\left(x\right)}\rightarrow\nonumber \\
        \rightarrow\frac{\delta S}{\delta\varphi\left(x\right)}
        \left[g,\frac{\delta}{\delta j}\right]=L\left(\partial\right)
        \frac{\delta}{\delta j\left(x\right)}+
        g\left(x\right)\int_{\lambda}\tilde{U}\left(\lambda\right)
        i\lambda\,\exp{\left(i\lambda\frac{\delta}{\delta j\left(x\right)}\right)}.
    \end{eqnarray}
This operator can be factored out from the path integral (\ref{ch2-eq1}) with respect to the primary field $\varphi$, after that, this functional integral transforms into the generating functional of total Green functions $\mathcal{Z}$. As a result, we obtain the functional equation in terms of functional derivatives, i.e. the Schwinger--Dyson equation:
    \begin{eqnarray}\label{ch3-eq3}
        \frac{\delta S}{\delta\varphi\left(x\right)}
        \left[g,\frac{\delta}{\delta j}\right]
        \mathcal{Z}\left[g,\,j\right]=
        j\left(x\right)\mathcal{Z}\left[g,j\right].
    \end{eqnarray}
The explicit form of equation (\ref{ch3-eq3}) is given by:
    \begin{eqnarray}\label{ch3-eq4}
        L\left(\partial\right)\frac{\delta\mathcal{Z}\left[g,j\right]}{\delta j\left(x\right)}+g\left(x\right)\int_{\lambda}
        \tilde{U}\left(\lambda\right)i\lambda
        \mathcal{Z}\left[g,j+i\lambda\delta_{\bullet x}\right]=
        j\left(x\right)\mathcal{Z}\left[g,j\right].
    \end{eqnarray}
In the equation (\ref{ch3-eq4}), the important notation is introduced: since the argument of the functional (for example, the source $j$) is a value with an undefined argument, it is common to denote this argument by a dot (for example $j_{\bullet}$, which is usually omitted for brevity). The argument $x$ in equation (\ref{ch3-eq4}) is the external parameter. In particular, if the argument of the functional is a sum of $a_{\bullet}+b_{\bullet x}$, then this notation means that the object $a$ with one argument is added to the object $b$ with two arguments. In this case the second argument of object $b$ is external and equal to $x$

Let us proceed from SD equation (\ref{ch3-eq4}) for generating functional of total Green functions $\mathcal{Z}$ to the similar equation for $\mathcal{S}$-matrix of theory. For this proceeding, all the substitutions which transform $\mathcal{Z}$ into $\mathcal{S}$ should be made. These substitutions are well-known from books \cite{kopbarsch} and \cite{vasil1998methods,vasil2004field}, for this reason, we briefly repeat the corresponding derivation. As the first step, we represent $\mathcal{Z}$ as the product of free theory functional $\mathcal{Z}_{0}$ and reminder functional $\mathcal{Z}_{1}$ (it does not have to be small) generated by the interaction action $S_{1}$. After obtaining the equation for $\mathcal{Z}_{1}$, one more substitution is performed -- the substitution of the functional arguments $j=\hat{L}\varphi$ where the external field $\varphi$ is denoted by the same symbol as the primary field in (\ref{ch2-eq1}) for functional $\mathcal{Z}$. The next equalities partially demonstrate this derivation:
    \begin{eqnarray}\label{ch3-eq5}
        \mathcal{Z}\left[g,j\right]=\mathcal{Z}_{0}\left[j\right]
        \mathcal{Z}_{1}
        \left[g,j\right],\quad\mathcal{Z}_{1}\left[g,j=\hat{L}\varphi\right]=
        \mathcal{S}\left[g,\varphi\right].
    \end{eqnarray}
After above calculations we obtain the SD equation for the $\mathcal{S}$-matrix:
    \begin{eqnarray}\label{ch3-eq6}
        \frac{\delta\mathcal{S}\left[g,\varphi\right]}{\delta\varphi\left(x\right)}+g\left(x\right)\int_{\lambda}\tilde{U}\left(\lambda\right)i\lambda\,e^{-\frac{\lambda^{2}}{2}G\left(0\right)+i\lambda\varphi\left(x\right)}\mathcal{S}\left[g,\varphi+i\lambda G_{\bullet x}\right]=0.
    \end{eqnarray}
According to expression (\ref{ch2-eq17}), the logarithm of the $\mathcal{S}$-matrix $\mathcal{G}$ is a generating functional of the amputated connected Green functions. It is also useful to write the SD equation for the generating functional $\mathcal{G}$:
    \begin{eqnarray}\label{ch3-eq7}
        \frac{\delta\mathcal{G}\left[g,\varphi\right]}
        {\delta\varphi\left(x\right)}+g\left(x\right)\int_{\lambda}\tilde{U}
        \left(\lambda\right)i\lambda\,e^{-\frac{\lambda^{2}}{2}G\left(0\right)+
        i\lambda\varphi\left(x\right)}
        e^{\mathcal{G}\left[g,\varphi+i\lambda
        G_{\bullet x}\right]-\mathcal{G}\left[g,\varphi\right]}=0.
    \end{eqnarray}
Equations (\ref{ch3-eq4}), (\ref{ch3-eq6}) and (\ref{ch3-eq7}) give us the full set of the SD equations which will be used to analyze different expressions for the $\mathcal{S}$-matrix in Efimov nonlocal and nonpolynomial QFT.

Now, let us focus on the derivation of another important equation in quantum field theory and statistical physics which is the functional Schr\"{o}dinger equation (also called the Tomonaga--Schwinger equation). To derive this equation, let us again consider the expression (\ref{ch2-eq1}) for the generating functional $\mathcal{Z}$. Let us take the functional derivative with respect to coupling constant (function) $g$. The result can be written as:
    \begin{eqnarray}\label{ch3-eq8}
        \int\mathcal{D}\left[\varphi\right]\frac{\delta e^{-S\left[g,\varphi\right]+\left(j\mid\varphi\right)}}{\delta g\left(x\right)}=-\int\mathcal{D}\left[\varphi\right]e^{-S\left[g,\varphi\right]+\left(j\mid\varphi\right)}\frac{\delta S_{1}\left[g,\varphi\right]}{\delta g\left(x\right)}.
    \end{eqnarray}
Then, we calculate the functional derivative with respect to $g$ of the interaction action $S_{1}$ under the sign of path integral with respect to $\varphi$. In this case, as in the derivation of the SD equation, for the interaction action $S_{1}$ we use Efimov representation (\ref{ch2-eq5}). Then we use source trick with $j$ again. The functional derivative with respect to the coupling constant $g$ of the interaction action $S_{1}$ becomes an operator in terms of the functional derivatives with respect to $j$ in a form of the functional translation operator. As a result of performed calculations, the following equation -- the functional Schr\"{o}dinger equation, is obtained:
    \begin{eqnarray}\label{ch3-eq9}
        \frac{\delta\mathcal{Z}\left[g,j\right]}{\delta g\left(x\right)}+
        \int_{\lambda}\tilde{U}\left(\lambda\right)\,
        \mathcal{Z}\left[g,j+i\lambda\delta_{\bullet x}\right]=0.
    \end{eqnarray}
The notations in the equation (\ref{ch3-eq9}) repeat the corresponding notations in the equation (\ref{ch3-eq4}).

As in the case of SD equation, the next step is to proceed from the functional Schr\"{o}dinger equation (\ref{ch3-eq9}) for the generating functional $\mathcal{Z}$ to the similar equation for the $\mathcal{S}$-matrix. The derivation repeat the expression (\ref{ch3-eq5}) and, as a result of performed transformations, the functional Schr\"{o}dinger equation for the $\mathcal{S}$-matrix is obtained:
    \begin{eqnarray}\label{ch3-eq10}
        \frac{\delta\mathcal{S}\left[g,\varphi\right]}{\delta g\left(x\right)}+\int_{\lambda}\tilde{U}\left(\lambda\right)
        e^{-\frac{\lambda^{2}}{2}G\left(0\right)+
        i\lambda\varphi\left(x\right)}
        \mathcal{S}\left[g,\varphi+i\lambda G_{\bullet x}\right]=0.
    \end{eqnarray}
Finally, recalling the relation (\ref{ch2-eq17}) between the $\mathcal{S}$-matrix and the generating functional of amputated connected Green functions $\mathcal{G}$, it is easy to obtain the functional Schr\"{o}dinger equation for the functional $\mathcal{G}$. This equation reads:
    \begin{eqnarray}\label{ch3-eq11}
        \frac{\delta\mathcal{G}\left[g,\varphi\right]}
        {\delta g\left(x\right)}+\int_{\lambda}\tilde{U}\left(\lambda\right)
        e^{-\frac{\lambda^{2}}{2}G\left(0\right)+
        i\lambda\varphi\left(x\right)}
        e^{\mathcal{G}\left[g,\varphi+i\lambda G_{\bullet x}\right]-
        \mathcal{G}\left[g,\varphi\right]}=0.
    \end{eqnarray}
Equations (\ref{ch3-eq9}), (\ref{ch3-eq10}) and (\ref{ch3-eq11}) give us the full set of functional Schr\"{o}dinger equations which we will use to analyze different expressions for the $\mathcal{S}$-matrix in Efimov nonlocal and nonpolynomial QFT equally with the similar set of the SD equations. We note that G.V. Efimov used the approach based on the functional Schr\"{o}dinger equation. It seems natural since the expressions (\ref{ch2-eq20})--(\ref{ch2-eq21}) are the expansion of the $\mathcal{S}$-matrix with respect to the coupling constant $g$. Nevertheless, an approach based on the SD equations allows us to shed the light on the types of the nonperturbative (not based on the expansion with respect to the coupling constant $g$) expressions for the $\mathcal{S}$-matrix.

\subsection{On Newton--Leibniz formula for functionals}

In this subsection, we derive the Newton--Leibniz (NL) formula for functionals. Detailed discussion of Newton--Leibniz formula derivation is interesting by the following reason: this derivation is similar to the analogical one in case of the FRG flow equations of Wilson--Polchinski or Wetterich--Morris \cite{kopbarsch}. Indeed, the introduction of any $\varLambda$-deformed functional with the subsequent differentiation with respect to scale $\varLambda$, and transformation of the obtained result into an expression in terms of functional derivatives with respect to the argument of functional are the key moments in the derivation of NL formula and different FRG flow equations. In other words, NL formula is the particular case of FRG flow equations.

Let us carry out the described derivation on the example of a functional $\mathcal{F}$. As first step, we add an increment $\varDelta_{\varLambda}$ depending on parameter $\varLambda$ to the argument $\varphi$ of functional $\mathcal{F}$. After this, we can rewrite the increment in terms of functional translation operator. Finally, by taking the derivative of the following expression with respect to scale $\varLambda$ and using the functional translation operator once again, we obtain the expression:
    \begin{eqnarray}\label{ch3-eq12}
        \mathcal{F}\left[\varphi+\varDelta_{\varLambda}\right]=e^{\left(\varDelta_{\varLambda}\left|\frac{\delta}{\delta\varphi}\right.\right)}\mathcal{F}\left[\varphi\right],\quad
        \frac{\partial\mathcal{F}\left[\varphi+\varDelta_{\varLambda}\right]}{\partial\varLambda}=\left(\frac{\partial\varDelta_{\varLambda}}{\partial\varLambda}\bigg|\frac{\delta\mathcal{F}\left[\varphi+\varDelta_{\varLambda}\right]}        {\delta\varphi}\right).
    \end{eqnarray}
Let us integrate this equation with respect to $\varLambda$ in the limits from $\varLambda_{1}$ to $\varLambda_{2}$. The result of integration reads:
    \begin{eqnarray}\label{ch3-eq13}
        \mathcal{F}\left[\varphi+\varDelta_{\varLambda_{2}}\right]-
        \mathcal{F}\left[\varphi+\varDelta_{\varLambda_{1}}\right]=
        \int\limits_{\varLambda_{1}}^{\varLambda_{2}}d\varLambda\int d^{D}x\,\frac{\partial\varDelta_{\varLambda}
        \left(x\right)}{\partial\varLambda}
        \frac{\delta\mathcal{F}\left[\varphi+\varDelta_{\varLambda}\right]}
        {\delta\varphi\left(x\right)}.
    \end{eqnarray}
To obtain the NL formula in a canonical form, the value $\varDelta_{\varLambda}$ should be chosen in the simplest form which is the product of $\varLambda$ and $\psi$, i.e. $\varDelta_{\varLambda}=\varLambda\psi$ (the dimensions of $\varphi$ and $\psi$ are coincide because the parameter $\varLambda$ is dimensionless). In this case, we obtain the following expression:
    \begin{eqnarray}\label{ch3-eq14}
        \mathcal{F}\left[\varphi+\varLambda_{2}\psi\right]-\mathcal{F}\left[\varphi+\varLambda_{1}\psi\right]=\int\limits_{\varLambda_{1}}^{\varLambda_{2}}d\varLambda\int d^{D}x\,\psi\left(x\right)\frac{\delta\mathcal{F}\left[\varphi+\varLambda\psi\right]}{\delta\varphi\left(x\right)}.
    \end{eqnarray}
Now let us choose $\varLambda_{1}=0$ and $\varLambda_{2}=1$. In this case we finally arrive to the canonical NL formula for functionals:
    \begin{eqnarray}\label{ch3-eq15}
        \mathcal{F}\left[\varphi+\psi\right]-\mathcal{F}\left[\varphi\right]=
        \int\limits_{0}^{1}d\varLambda\int d^{D}x\,\psi\left(x\right)
        \frac{\delta\mathcal{F}\left[\varphi+\varLambda\psi\right]}
        {\delta\varphi\left(x\right)}.
    \end{eqnarray}
The obtained expressions reconstruct the functional by its first functional derivative. Of course, for the self-consistency of obtained construction, several additional conditions have to be satisfied. In particular, the second functional derivative of the functional $\mathcal{F}$ has to be independent of the variation order. These conditions are satisfied in all the constructions considered in the present paper.

\subsection{Schwinger--Dyson and functional Schr\"{o}dinger equations in completely integral form}

The Schwinger--Dyson equations (\ref{ch3-eq6})--(\ref{ch3-eq7}), as the functional Schr\"{o}dinger equations (\ref{ch3-eq10})--(\ref{ch3-eq11}), contain the functionals $\mathcal{S}$ and $\mathcal{G}$ with shifted arguments $\varphi+i\lambda G_{\bullet x}$ and first functional derivatives of functionals $\mathcal{S}$ and $\mathcal{G}$ with normal arguments. It is convenient to rewrite these equations using the Newton--Leibniz formula from the previous subsection. In this case, we obtain the SD equations and the functional Schr\"{o}dinger equations in completely integral form. The completely integral form of the SD equation for $\mathcal{S}$-matrix can be written as:
    \begin{eqnarray}\label{ch3-eq16}
        \mathcal{S}\left[g,\varphi+\varLambda_{1}\psi\right]-\mathcal{S}
        \left[g,\varphi+\varLambda_{2}\psi\right]=
        \int\limits_{\varLambda_{1}}^{\varLambda_{2}}d\varLambda\int d^{D}x\,\psi\left(x\right)g\left(x\right)\times\nonumber\\
        \times\int_{\lambda}\tilde{U}\left(\lambda\right)
        i\lambda\,e^{-\frac{\lambda^{2}}{2}G\left(0\right)+
        i\lambda\left[\varphi\left(x\right)+
        \varLambda\psi\left(x\right)\right]}
        \mathcal{S}\left[g,\varphi+\varLambda\psi+
        i\lambda G_{\bullet x}\right].
    \end{eqnarray}
Similarly, the completely integral form of the SD equation for generating functional $\mathcal{G}$ is given by:
    \begin{eqnarray}\label{ch3-eq17}
        \mathcal{G}\left[g,\varphi+\varLambda_{1}\psi\right]-
        \mathcal{G}\left[g,\varphi+\varLambda_{2}\psi\right]=
        \int\limits_{\varLambda_{1}}^{\varLambda_{2}}d\varLambda\int d^{D}x\,\psi\left(x\right)g\left(x\right)\times\nonumber\\
        \times\int_{\lambda}
        \tilde{U}\left(\lambda\right)i\lambda\, e^{-\frac{\lambda^{2}}{2}G\left(0\right)+
        i\lambda\left[\varphi\left(x\right)+ \varLambda\psi\left(x\right)\right]}
        e^{+\mathcal{G}\left[g,\varphi+\varLambda\psi+
        i\lambda G_{\bullet x}\right]-
        \mathcal{G}\left[g,\varphi+\varLambda\psi\right]}.
    \end{eqnarray}
Then, the completely integral form of the functional Schr\"{o}dinger equation for $\mathcal{S}$-matrix can be written as:
    \begin{eqnarray}\label{ch3-eq18}
        \mathcal{S}\left[g+\varLambda_{1}h,\varphi\right]-\mathcal{S}\left[g+\varLambda_{2}h,\varphi\right]=\int\limits_{\varLambda_{1}}^{\varLambda_{2}}d\varLambda\int d^{D}x\,h\left(x\right)\times\nonumber\\ \times\int_{\lambda}\tilde{U}\left(\lambda\right)e^{-\frac{\lambda^{2}}{2}G\left(0\right)+i\lambda\varphi\left(x\right)}\mathcal{S}\left[g+\varLambda h,\varphi+i\lambda G_{\bullet x}\right]\,.
    \end{eqnarray}
Finally, the completely integral form of the functional Schr\"{o}dinger equation for the generating functional $\mathcal{G}$ reads:
    \begin{eqnarray}\label{ch3-eq19}
        \mathcal{G}\left[g+\varLambda_{1}h,\varphi\right]-\mathcal{G}\left[g+\varLambda_{2}h,\varphi\right]=\int\limits_{\varLambda_{1}}^{\varLambda_{2}}d\varLambda\int d^{D}x\,h\left(x\right)\times\nonumber\\ \times\int_{\lambda}\tilde{U}\left(\lambda\right)e^{-\frac{\lambda^{2}}{2}G\left(0\right)+i\lambda\varphi\left(x\right)+\mathcal{G}\left[g+\varLambda h,\varphi+i\lambda G_{\bullet x}\right]-\mathcal{G}\left[g+\varLambda h,\varphi\right]}\,.
    \end{eqnarray}
In the next subsections, we focus on the integral forms of (\ref{ch3-eq16})--(\ref{ch3-eq17}). Instead of analyzing the dynamics within the coupling constant $g$, i.e. instead of discussing the functional Schr\"{o}dinger equations, we assume that $g$ is given (fixed in some form). Discussion of the equations (\ref{ch3-eq18})--(\ref{ch3-eq19}) can be a subject for the future publication.

\subsection{Solution of Schwinger--Dyson equation: infinite field limit}

In this subsection, we obtain the solutions of the Schwinger--Dyson equations (\ref{ch3-eq16})--(\ref{ch3-eq17}) in the limit of the infinite field configurations. As the first step we set the value $\varphi$ to be equal to zero. This step should not lead to confusion because in the title we assume the field $\psi$ is infinite. As a result of the first step, the equation (\ref{ch3-eq16}) for the $\mathcal{S}$-matrix transforms to the following form:
    \begin{eqnarray}\label{ch3-eq20}
        \mathcal{S}\left[g,\varLambda_{1}\psi\right]-\mathcal{S}\left[g,\varLambda_{2}\psi\right]=\int\limits_{\varLambda_{1}}^{\varLambda_{2}}d\varLambda\int d^{D}x\,\psi\left(x\right)g\left(x\right)\times\nonumber\\ \times\int_{\lambda}\tilde{U}\left(\lambda\right)i\lambda\,e^{-\frac{\lambda^{2}}{2}G\left(0\right)+i\lambda\varLambda\psi\left(x\right)}\mathcal{S}\left[g,\varLambda\psi+i\lambda G_{\bullet x}\right].
    \end{eqnarray}
The equation (\ref{ch3-eq17}) for the generating functional $\mathcal{G}$ can be written as:
    \begin{eqnarray}\label{ch3-eq21}
        \mathcal{G}\left[g,\varLambda_{1}\psi\right]-\mathcal{G}\left[g,\varLambda_{2}\psi\right]=\int\limits_{\varLambda_{1}}^{\varLambda_{2}}d\varLambda\int d^{D}x\,\psi\left(x\right)g\left(x\right)\times\nonumber\\ \times\int_{\lambda}\tilde{U}\left(\lambda\right)i\lambda\,e^{-\frac{\lambda^{2}}{2}G\left(0\right)+i\lambda\varLambda\psi\left(x\right)+\mathcal{G}\left[g,\varLambda\psi+i\lambda G_{\bullet x}\right]-\mathcal{G}\left[g,\varLambda\psi\right]}.\label{1limsol2}
    \end{eqnarray}
We note that the equations (\ref{ch3-eq20})--(\ref{ch3-eq21}) are exact. As the second step we consider infinite field configurations $\psi\rightarrow\infty$. In the framework of this approximation, it is convenient to analyze the equation (\ref{ch3-eq21}). Now, this equation is a simple equality with known right-hand side:
    \begin{eqnarray}\label{ch3-eq22}
        \mathcal{G}\left[g,\varLambda_{1}\psi\right]-\mathcal{G}\left[g,\varLambda_{2}\psi\right]\approx\int\limits_{\varLambda_{1}}^{\varLambda_{2}}d\varLambda\int d^{D}x\,
        \psi\left(x\right)g\left(x\right)\int_{\lambda}\tilde{U}\left(\lambda\right)i\lambda\,e^{-\frac{\lambda^{2}}{2}G\left(0\right)+i\lambda\varLambda\psi\left(x\right)}.
    \end{eqnarray}
The integral with respect to $\varLambda$ in the limits from $\varLambda_{1}$ to $\varLambda_{2}$ can be calculated explicitly. After integration, we obtain:
    \begin{eqnarray}\label{ch3-eq23}
        \mathcal{G}\left[g,\varLambda_{1}\psi\right]-\mathcal{G}\left[g,\varLambda_{2}\psi\right]\approx\int d^{D}x\,
        g\left(x\right)\int_{\lambda}\tilde{U} \left(\lambda\right)e^{-\frac{\lambda^{2}}{2}G\left(0\right)}\left[e^{i\lambda\varLambda_{2}\psi\left(x\right)}-e^{i\lambda\varLambda_{1}\psi\left(x\right)}\right].
    \end{eqnarray}
The equation (\ref{ch3-eq23}) means that in the infinite field configurations limit the generating functional $\mathcal{G}$ almost transforms into the interaction action $S_{1}$ but with a certain {\bf\emph{modulation}} of the Fourier image $\tilde{U}$ of the interaction Lagrangian $U$:
    \begin{eqnarray}\label{ch3-eq24}
        \mathcal{G}\left[g,\varphi\right]=\mathcal{C}-\int d^{D}x\,g\left(x\right)\int_{\lambda}
        \tilde{U}\left(\lambda\right
        )e^{-\frac{\lambda^{2}}{2}G\left(0\right)+
        i\lambda\varphi\left(x\right)},\quad
        \mathcal{S}\left[g,\varphi\right]=
        e^{\mathcal{G}\left[g,\varphi\right]}.
    \end{eqnarray}    
Therefore, the expression (\ref{ch3-eq24}) gives the solutions for the desired functionals $\mathcal{S}$ and $\mathcal{G}$ in the infinite field limit.

\subsection{Solution of Schwinger--Dyson equation: zero field limit}

In the following and further subsections, we obtain and analyze the solutions of the approximated Schwinger--Dyson equations from which the solution in the limit of zero field configurations follows. Although we focus on the sine-Gordon theory in both subsections, obtained conclusions have a sufficiently general nature. For sine-Gordon theory, the Fourier image $\tilde{U}$ of the interaction Lagrangian $U$ is given by ($\alpha$ is the constant parameter):
    \begin{eqnarray}\label{ch3-eq25}
        \tilde{U}\left(\lambda\right)=
        \pi\left[\delta\left(\lambda-\alpha\right)+
        \delta\left(\lambda+\alpha\right)\right].
    \end{eqnarray}
The approximated SD equations are obtained from the exact equations (\ref{ch3-eq20})--(\ref{ch3-eq21}) if we set the field $\psi$ to be equal to the given and fixed field $\psi_{0}$, and the value of $\varLambda$ to be equal to fixed value $\varLambda_{0}$. The approximated equation for the $\mathcal{S}$-matrix has the following form:
    \begin{eqnarray}\label{ch3-eq26}
        \mathcal{S}\left[g,\varLambda_{1}\psi\right]-\mathcal{S}\left[g,\varLambda_{2}\psi\right]\approx\int\limits_{\varLambda_{1}}^{\varLambda_{2}}d\varLambda\int d^{D}x\,\psi\left(x\right)
        \bar{g}\left(x\right)\times\nonumber\\ \times\frac{i\alpha}{2}\Big\{e^{i\alpha\varLambda\psi\left(x\right)}\mathcal{S}\left[g,\varLambda_{0}\psi_{0}+i\alpha G_{\bullet x}\right]-
        e^{-i\alpha\varLambda\psi\left(x\right)}\mathcal{S}
        \left[g,\varLambda_{0}\psi_{0}-i\alpha G_{\bullet x}\right]\Big\}.
    \end{eqnarray}
The approximated equation for the generating functional $\mathcal{G}$ reads:
    \begin{eqnarray}\label{ch3-eq27}
        \mathcal{G}\left[g,\varLambda_{1}\psi\right]-\mathcal{G}\left[g,\varLambda_{2}\psi\right]\approx\int\limits_{\varLambda_{1}}^{\varLambda_{2}}d\varLambda\int d^{D}x\,\psi\left(x\right)
        \bar{g}\left(x\right)e^{-\mathcal{G}\left[g,\varLambda_{0}\psi_{0}\right]}\times\nonumber\\ \times\frac{i\alpha}{2}\Big\{e^{i\alpha\varLambda
        \psi\left(x\right)+
        \mathcal{G} \left[g,\varLambda_{0}\psi_{0}+i\alpha G_{\bullet x}\right]}-
        e^{-i\alpha\varLambda\psi\left(x\right)+\mathcal{G}\left[g,\varLambda_{0}\psi_{0}-i\alpha G_{\bullet x}\right]}\Big\}.
    \end{eqnarray}
Here we introduced the compact notation $\bar{g}$ for coupling constant, which absorbs additional exponential constant. This is an example of renormalization. The solution of the obtained equations (\ref{ch3-eq26})--(\ref{ch3-eq27}) is the main goal of current and next subsections.

Both obtained equations can be simplified because the integral with respect to $\varLambda$ in the limits from $\varLambda_{1}$ to $\varLambda_{2}$ can be calculated explicitly. As a result, we arrive to the following equation for the $\mathcal{S}$-matrix:
    \begin{eqnarray}\label{ch3-eq28}
        \mathcal{S}\left[g,\varLambda_{1}\psi\right]-\mathcal{S} \left[g,\varLambda_{2}\psi\right]\approx\frac{1}{2}\int d^{D}x\,\bar{g}\left(x\right)\times\nonumber\\ \times\Big\{\left[e^{i\alpha\varLambda_{2}\psi\left(x\right)}-e^{i\alpha\varLambda_{1}\psi\left(x\right)}\right]\mathcal{S}\left[g,\varLambda_{0}\psi_{0}+i\alpha G_{\bullet x}\right]+\nonumber\\
        +\left[e^{-i\alpha\varLambda_{2}\psi\left(x\right)}-e^{-i\alpha\varLambda_{1}\psi\left(x\right)}\right]\mathcal{S}\left[g,\varLambda_{0}\psi_{0}-i\alpha G_{\bullet x}\right]\Big\}.
    \end{eqnarray}
A similar equation for the generating functional $\mathcal{G}$ can be written as:
    \begin{eqnarray}\label{ch3-eq29}
        \mathcal{G}\left[g,\varLambda_{1}\psi\right]-\mathcal{G}\left[g,\varLambda_{2}\psi\right]\approx\frac{1}{2}\int d^{D}x\,\bar{g}\left(x\right)e^{-\mathcal{G}
        \left[g,\varLambda_{0}\psi_{0}\right]}\times\nonumber\\ \times\Big\{\left[e^{i\alpha\varLambda_{2}\psi\left(x\right)}-e^{i\alpha\varLambda_{1}\psi\left(x\right)}\right]e^{\mathcal{G}\left[g,\varLambda_{0}\psi_{0}+i\alpha G_{\bullet x}\right]}+\nonumber\\
        +\left[e^{-i\alpha\varLambda_{2}\psi\left(x\right)}-e^{-i\alpha\varLambda_{1}\psi\left(x\right)}\right]e^{\mathcal{G}\left[g,\varLambda_{0}\psi_{0}-i\alpha G_{\bullet x}\right]}\Big\}.
    \end{eqnarray}
For the further discussions, it is convenient to define the field $\varphi_{0}=\varLambda_{0}\psi_{0}$. Also, because the coupling constant $g$ is a given and fixed function, from now on the dependence of functionals on $g$ will be omitted.

Because the equations (\ref{ch3-eq28})--(\ref{ch3-eq29}) are equalities with known right-hand side, corresponding solutions are found automatically. The solution for the $\mathcal{S}$-matrix of theory has the following form (the index $\pm$ denotes a corresponding shift of argument):
    \begin{eqnarray}\label{ch3-eq30}
        \mathcal{S}\left[\varphi\right]=\mathcal{C}-\frac{1}{2}\int d^{D}x\,\bar{g}\left(x\right)\Big\{e^{i\alpha\varphi\left(x\right)} \mathcal{S}_{+}\left[\varphi_{0},x\right]+
        e^{-i\alpha\varphi\left(x\right)}
        \mathcal{S}_{-}\left[\varphi_{0},x\right]\Big\}.
    \end{eqnarray}
Also, the solution for the generating functional $\mathcal{G}$ can be written as:
    \begin{eqnarray}\label{ch3-eq31}
        \mathcal{G}\left[\varphi\right]=\tilde{\mathcal{C}}-\frac{1}{2}\int d^{D}x\,\bar{g}\left(x\right)e^{-\mathcal{G}\left[\varphi_{0}\right]}
        \Big\{e^{i\alpha\varphi\left(x\right)+\mathcal{G}_{+}
        \left[\varphi_{0},x\right]}+
        e^{-i\alpha\varphi\left(x\right)+
        \mathcal{G}_{-}\left[\varphi_{0},x\right]}\Big\}.
    \end{eqnarray}
The expressions (\ref{ch3-eq30})--(\ref{ch3-eq31}) are the solutions of the approximated SD equations. However, these solutions are integral equations by themselves, i.e. contains themselves at the field values $\varphi=\varphi_{0}$ and $\varphi=\varphi_{0}\pm i\alpha G_{\bullet x}$ in the right-hand side. Further, we find the closed equations for corresponding values, i.e. self-consistency equations. We solve them approximately and, in this way, obtain the final expressions for the functionals $\mathcal{S}$ and $\mathcal{G}$.

\subsection{Solution of self-consistency equations and analysis of the results}

Let us consider the solution for the $\mathcal{S}$-matrix (\ref{ch3-eq30}). To obtain closed integral equations for values $\mathcal{S}_{+}$ and $\mathcal{S}_{-}$, we substitute the field configurations $\varphi=\varphi_{0}\pm i\alpha G_{\bullet z}$ (external spatial argument is $z$) to expression (\ref{ch3-eq30}). Here we note that for explicit calculations, it is possible to use the well-known expression for the massive propagator \cite{vasil1998methods,vasil2004field} in order to improve the convergence of integrals at infinity. Also, we remind that the propagator is regularized at zero $x$ due to the form factor. As a result, we obtain two coupled equations for $\mathcal{S}_{+}$ and $\mathcal{S}_{-}$:
        \begin{eqnarray}\label{ch3-eq32}
        \mathcal{S}_{+}\left[\varphi_{0},z\right]=\mathcal{C}-\frac{1}{2}\int d^{D}x\,\bar{g}\left(x\right)\times\nonumber\\
        \times\Big\{e^{i\alpha\varphi_{0}\left(x\right)-\alpha^{2}G\left(x-z\right)}\mathcal{S}_{+}\left[\varphi_{0},x\right]+
        e^{-i\alpha\varphi_{0}\left(x\right)+\alpha^{2}G\left(x-z\right)}\mathcal{S}_{-}\left[\varphi_{0},x\right]\Big\},\nonumber\\\mathcal{S}_{-}\left[\varphi_{0},z\right]=\mathcal{C}-\frac{1}{2}\int d^{D}x\,\bar{g}\left(x\right)\times\nonumber\\
        \times\Big\{e^{i\alpha\varphi_{0}\left(x\right)+\alpha^{2}G\left(x-z\right)}\mathcal{S}_{+}
        \left[\varphi_{0},x\right]+
        e^{-i\alpha\varphi_{0}\left(x\right)-
        \alpha^{2}G\left(x-z\right)}\mathcal{S}_{-}
        \left[\varphi_{0},x\right]\Big\}.
    \end{eqnarray}
These equations (\ref{ch3-eq32}) are very complicated mathematical objects because $\mathcal{S}_{+}$ and $\mathcal{S}_{-}$ depend on the arbitrary field $\varphi_{0}$ functionally. In the present paper we limit ourselves to the simplest case: assume that $\varphi_{0}=\mathrm{const}$. As it will be seen from the results for $\mathcal{S}_{+}$ and $\mathcal{S}_{-}$, this ansatz is enough to find a variety of interesting conclusions. In the case of constant fields, we obtain two coupled equations but now $\mathcal{S}_{+}$ and $\mathcal{S}_{-}$ are functions (the arguments of the corresponding functions are in the parentheses):
    \begin{eqnarray}\label{ch3-eq33}
        \mathcal{S}_{+}\left(\varphi_{0},z\right)=\mathcal{C}-\frac{1}{2}\int d^{D}x\,\bar{g}\left(x\right)\times\nonumber\\
        \times\Big\{e^{i\alpha\varphi_{0}-\alpha^{2}G\left(x-z\right)}\mathcal{S}_{+}\left(\varphi_{0},x\right)+
        e^{-i\alpha\varphi_{0}+\alpha^{2}G\left(x-z\right)}\mathcal{S}_{-}\left(\varphi_{0},x\right)\Big\},\nonumber\\\mathcal{S}_{-}\left(\varphi_{0},z\right)=\mathcal{C}-\frac{1}{2}\int d^{D}x\,\bar{g}\left(x\right)\times\nonumber\\
        \times\Big\{e^{i\alpha\varphi_{0}+\alpha^{2}G\left(x-z\right)}\mathcal{S}_{+}\left(\varphi_{0},x\right)+
        e^{-i\alpha\varphi_{0}-\alpha^{2}G\left(x-z\right)}
        \mathcal{S}_{-}\left(\varphi_{0},x\right)\Big\}.
    \end{eqnarray}
To find the analytical solution of the system of equations (\ref{ch3-eq33}), let us recall the example from the theory of superconductivity. Equations of this type are usual for the latter. To obtain the analytical estimation, the step ansatz for solution is used (we can always improve it in the framework of the separable approximation and other). This ansatz is well justified if the functions characterizing the system are sign-constant. 

In the light of all the above, the system of equations (\ref{ch3-eq33}) can be replaced by the simplified (approximated) one:
    \begin{eqnarray}\label{ch3-eq34}
        \mathcal{S}_{+}\left(\varphi_{0}\right)=\mathcal{C}-e^{i\alpha\varphi_{0}}I_{-}\mathcal{S}_{+}\left(\varphi_{0}\right)-e^{-i\alpha\varphi_{0}}I_{+}\mathcal{S}_{-}\left(\varphi_{0}\right),\nonumber\\ \mathcal{S}_{-}\left(\varphi_{0}\right)=\mathcal{C}-e^{i\alpha\varphi_{0}}I_{+}\mathcal{S}_{+}\left(\varphi_{0}\right)-e^{-i\alpha\varphi_{0}}I_{-}\mathcal{S}_{-}\left(\varphi_{0}\right).
    \end{eqnarray}
In the system of equations (\ref{ch3-eq34}) the following notations are introduced:
    \begin{eqnarray}\label{ch3-eq35}
        I_{+}=\frac{1}{2}\int d^{D}x\,\bar{g}\left(x\right)
        e^{\alpha^{2}G\left(x\right)},\quad 
        I_{-}=\frac{1}{2}\int d^{D}x\,\bar{g}\left(x\right)
        e^{-\alpha^{2}G\left(x\right)}.
    \end{eqnarray}
The solution of (\ref{ch3-eq34})--(\ref{ch3-eq35}) is easily to found in the analytical form. Thus, we arrive at the following expression for for the desired functions $\mathcal{S}_{+}$ and $\mathcal{S}_{-}$:
    \begin{eqnarray}\label{ch3-eq36}
        \mathcal{S}_{+}\left(\varphi_{0}\right)=\mathcal{C}\,\frac{e^{-i\alpha\varphi_{0}}\left(I_{+}-I_{-}\right)-1}{I_{+}^{2}-I_{-}^{2}-1-2I_{-}\cos\left(\alpha\varphi_{0}\right)}\,,\nonumber\\ \mathcal{S}_{-}\left(\varphi_{0}\right)=\mathcal{C}\,\frac{e^{i\alpha\varphi_{0}}\left(I_{+}-I_{-}\right)-1}{I_{+}^{2}-I_{-}^{2}-1-2I_{-}\cos\left(\alpha\varphi_{0}\right)}\,.
    \end{eqnarray}
These expressions are interesting to analyze. First of all, generally, $I_{+}\gg I_{-}$. Therefore, the denominator is never equal to zero. This means that the system is not quantum trivial. This is the main conclusion from the expression (\ref{ch3-eq36}). Secondly, the expression (\ref{ch3-eq36}) demonstrates the dependencies which are hard to see from, for example, the expressions (\ref{ch2-eq20})--(\ref{ch2-eq21}) for the $\mathcal{S}$-matrix of the theory in terms of the Taylor series with respect to the interaction (coupling) constant $g$. Thirdly, the expressions similar to (\ref{ch3-eq36}) can be used as an initial point for a new expansion of the $\mathcal{S}$-matrix. This expansion is follows directly from the Schwinger--Dyson equation, for example in the form (\ref{ch3-eq20}). This statement is valid not only for sine-Gordon theory but also for a wide class of theories in general. A detailed construction of such an expansion is a very interesting question but this question should be the subject of an independent research and publication.

Let us suppose that the expressions (\ref{ch3-eq30})--(\ref{ch3-eq36}) are known. These expressions should be used to start an iteration procedure for the solution of the functional equation (\ref{ch3-eq20}) in the following form:
    \begin{eqnarray}\label{ch3-eq37}
        \mathcal{S}^{\left(M+1\right)}
        \left[g,\varLambda_{1}\psi\right]-
        \mathcal{S}^{\left(M+1\right)}
        \left[g,\varLambda_{2}\psi\right]=
        \int\limits_{\varLambda_{1}}^{\varLambda_{2}}d\varLambda
        \int d^{D}x\,\psi\left(x\right)g\left(x\right)
        \times\nonumber\\
        \times\int_{\lambda}
        \tilde{U}\left(\lambda\right)i\lambda\,
        e^{-\frac{\lambda^{2}}{2}G\left(0\right)+
        i\lambda\varLambda\psi\left(x\right)}
        \mathcal{S}^{\left(M\right)}\left[g,\varLambda\psi+
        i\lambda G_{\bullet x}\right].
    \end{eqnarray}
The left-hand side of (\ref{ch3-eq37}) is calculated at the $(M+1)$-th step via the known right-hand side of (\ref{ch3-eq37}) which is calculated at the $M$-th step. The expressions (\ref{ch3-eq30})--(\ref{ch3-eq36}) are used at the first step of the iteration as a known right-hand side of (\ref{ch3-eq37}).

At this moment let us go back to the expressions (\ref{ch3-eq30})--(\ref{ch3-eq36}) again. If they are known, it is possible to find, as an example, $n$-particle Green functions which are generated by the $\mathcal{S}$-matrix of the theory. Here we generalize the definition of the corresponding Green functions: we define these functions at the arbitrary (in general, non-zero) value of constant field $\varphi_{0}=\mathrm{const}$ (the dependence on the coupling constant $g$ is omitted):
    \begin{eqnarray}\label{ch3-eq38}
        \mathcal{S}^{(n)}\left(\varphi_{0};x_{1},\ldots,x_{n}\right)=\frac{\delta^{n}\mathcal{S}\left[g,\varphi\right]}{\delta\varphi\left(x_{1}\right)\ldots\delta\varphi\left(x_{n}\right)}\bigg|_{\varphi=0,\,\varphi_{0}=\mathrm{const}}.
    \end{eqnarray}
We write down the corresponding Green functions for $n=0, 1$ and $2$. In the case of $n=0$ we obtain the vacuum expectation value $\mathcal{S}^{(0)}$. This value is not determined from the Schwinger--Dyson equation but it can be used instead of a functional $\mathcal{C}$ which in general depends on $\varphi_{0}$:
    \begin{eqnarray}\label{ch3-eq39}
        \mathcal{S}^{(0)}\left(\varphi_{0}\right)=\mathcal{C}\left(\varphi_{0}\right)\left[1-\zeta\frac{\left(I_{+}-I_{-}\right)\cos\left(\alpha\varphi_{0}\right)-1}{I_{+}^{2}-I_{-}^{2}-1-2I_{-}\cos\left(\alpha\varphi_{0}\right)}\right],\quad
        \zeta=\int d^{D}x\,\bar{g}\left(x\right).
    \end{eqnarray}
For one-particle Green function $\mathcal{S}^{(1)}$ we have the following expression:    
    \begin{eqnarray}\label{ch3-eq40}
        \mathcal{S}^{(1)}\left(\varphi_{0};x\right)=
        -\frac{\mathcal{S}^{(0)}\left(\varphi_{0}\right)}
        {\mathcal{N}\left(\varphi_{0}\right)}\,
        \bar{g}\left(x\right)\alpha\left(I_{+}-I_{-}\right)
        \sin\left(\alpha\varphi_{0}\right),\nonumber\\
        \mathcal{N}\left(\varphi_{0}\right)=
        I_{+}^{2}-I_{-}^{2}-1-2I_{-}
        \cos\left(\alpha\varphi_{0}\right)-
        \zeta\left[\left(I_{+}-I_{-}\right)
        \cos\left(\alpha\varphi_{0}\right)-1\right].
    \end{eqnarray}
We note that at $\varphi_{0}=0$ the one-particle Green function $\mathcal{S}^{(1)}$ vanishes identically, as it should be in the case of an even interaction Lagrangian $U$ with respect to the field. Also, we note that within the framework of the approximation under consideration it is possible to neglect the coupling constant $\bar{g}$ dependence on the space variable $x$. Therefore, the function $\mathcal{S}^{(1)}$ does not depend on $x$. 

Finally, the two-particle Green function can be written as:    
    \begin{eqnarray}\label{ch3-eq41}
        \mathcal{S}^{(2)}\left(\varphi_{0};x,y\right)=
        \frac{\mathcal{S}^{(0)}\left(\varphi_{0}\right)}
        {\mathcal{N}\left(\varphi_{0}\right)}\,\bar{g}
        \left(x\right)\alpha^{2}\delta^{\left(D\right)}
        \left(x-y\right)\left[\left(I_{+}-I_{-}\right)
        \cos\left(\alpha\varphi_{0}\right)-1\right].
    \end{eqnarray}
Our approximation allows us to neglect coupling constant $\bar{g}$ dependence on space variable $x$, thus, the two-particle Green function $\mathcal{S}^{(2)}$ is the translationally invariant quantity. The latter, however, contains the integral $\zeta$ of the coupling constant $\bar{g}$, which indicates an implicit breaking of the translational invariance.

This concludes the section devoted to the derivation and solution of the functional Schwinger--Dyson and Schr\"{o}dinger equations in Efimov representation. For the sake of completeness of our discussion of nonlocal and nonpolynomial QFT, in the next section we focus on the FRG flow and the holographic renormalization group equations (functional Hamilton--Jacobi equation can be found in the papers \cite{de2000holographic,verlinde2000rg,fukuma2003holographic,lizana2016holographic,akhmedov2003notes,akhmedov2011hints,heemskerk2011holographic}). For instance, classical constructions with extra dimensions have already appeared above in the expression (\ref{ch2-eq20})--(\ref{ch2-eq21}) for the $\mathcal{S}$-matrix of the theory. For this reason, ``RG finale'' gives a holistic meaning to Efimov nonlocal and nonpolynomial QFT.

\section{Functional \& holographic RG}\label{ch4}

\subsection{Wilson--Polchinski FRG flow equations}

An abstract functional renormalization group flow equation can be obtained by a $\varLambda$-deformation of the $\mathcal{S}$-matrix of the theory (by the introduction to the propagator $G$ in expressions (\ref{ch2-eq17}) and (\ref{ch2-eq18}) of an additional field $\varLambda$ which can have a various nature: external field, momentum-depending mass, etc) with subsequent functional differentiation of the representation (\ref{ch2-eq18}) for the $\mathcal{S}$-matrix with respect to the field $\varLambda$ (see, for instance, \cite{kopbarsch,wipf2012statistical,rosten2012fundamentals,igarashi2009realization}). Because of the interaction action $S_{1}$ does not depend on the field $\varLambda$ the result of this differentiation can be easily represented in terms of the second functional derivative of  $\mathcal{S}$ with respect to the field $\varphi$. The following expression illustrates this:
    \begin{eqnarray}\label{ch4-eq1}
        \frac{\delta\mathcal{S}\left[\varLambda,g,\varphi\right]}{\delta\varLambda}=
        \frac{1}{2}\mathrm{Tr}\left\{\frac{\delta\mathbf{G}\left[\varLambda\right]}{\delta\varLambda}\,
        \frac{\delta^{\otimes2}\mathcal{S}\left[\varLambda,g,\varphi\right]}
        {\delta\varphi^{\otimes2}}\right\}.
    \end{eqnarray}
The abstract equation (\ref{ch4-eq1}) can be rewritten in the coordinate representation:
    \begin{eqnarray}\label{ch4-eq2}
        \frac{\delta\mathcal{S}\left[\varLambda,g,\varphi\right]}
        {\delta\varLambda\left(z\right)}=
        \frac{1}{2}\int d^{D}x_{1}\int d^{D}x_{2}
        \frac{\delta G\left[\varLambda\right]
        \left(x_{1},x_{2}\right)}{\delta\varLambda\left(z\right)}\,
        \frac{\delta^{2}\mathcal{S}
        \left[\varLambda,g,\varphi\right]}
        {\delta\varphi\left(x_{1}\right)
        \delta\varphi\left(x_{2}\right)}.
    \end{eqnarray}
Starting from the equations (\ref{ch4-eq1})--(\ref{ch4-eq2}), we can obtain analogous equations for the generating functional of the amputated connected Green functions $\mathcal{G}$. The abstract FRG flow equation of the functional $\mathcal{G}$ is given by:
    \begin{eqnarray}\label{ch4-eq3}
        \frac{\delta\mathcal{G}\left[\varLambda,g,\varphi\right]}
        {\delta\varLambda}=\frac{1}{2}\mathrm{Tr}
        \left\{\frac{\delta\mathbf{G}\left[\varLambda\right]}
        {\delta\varLambda}\,\frac{\delta^{\otimes2}
        \mathcal{G}\left[\varLambda,g,\varphi\right]}
        {\delta\varphi^{\otimes2}}\right\}+\nonumber\\
        +\frac{1}{2}\left(\frac{\delta\mathcal{G}
        \left[\varLambda,g,\varphi\right]}
        {\delta\varphi}\bigg|\frac{\delta\mathbf{G}
        \left[\varLambda\right]}{\delta\varLambda}\,
        \frac{\delta\mathcal{G}\left[\varLambda,g,\varphi\right]}
        {\delta\varphi}\right).
    \end{eqnarray}
Finally, the abstract equation (\ref{ch4-eq3}) in the coordinate representation reads:
    \begin{eqnarray}\label{ch4-eq4}
        \frac{\delta\mathcal{G}\left[\varLambda,g,\varphi\right]}
        {\delta\varLambda\left(z\right)}=\frac{1}{2}\int d^{D}x_{1}
        \int d^{D}x_{2}\,\frac{\delta G\left[\varLambda\right]\left(x_{1},x_{2}\right)}
        {\delta\varLambda\left(z\right)}\times\nonumber\\
        \times\left\{\frac{\delta^{2}
        \mathcal{G}\left[\varLambda,g,\varphi\right]}
        {\delta\varphi\left(x_{1}\right)\delta\varphi\left(x_{2}\right)}+
        \frac{\delta\mathcal{G}\left[\varLambda,g,\varphi\right]}
        {\delta\varphi\left(x_{1}\right)}\frac{\delta\mathcal{G}
        \left[\varLambda,g,\varphi\right]}
        {\delta\varphi\left(x_{2}\right)}\right\}.
    \end{eqnarray}
The former and the latter equations (\ref{ch4-eq3})--(\ref{ch4-eq4}) are called the Wilson--Polchinski FRG flow equations (in the abstract form and in the coordinate representation, correspondingly). The equations (\ref{ch4-eq1})--(\ref{ch4-eq2}) for the functional $\mathcal{S}$ and also (\ref{ch4-eq3})--(\ref{ch4-eq4}) for the functional $\mathcal{G}$ are the natural complement to the Schwinger--Dyson and Schr\"{o}dinger ones. This can be seen at least from the fact that in the derivation of the Schr\"{o}dinger equation the explicit form of the interaction action $S_{1}$ plays the main role. However here the free theory action $S_{0}$ is the main. At the same time the solutions of all mentioned equation do not contradict one another. 

Unfortunately, the expansion (\ref{ch2-eq20})--(\ref{ch2-eq21}) is relatively easy to obtain by a direct calculation of a functional integral for a corresponding $\mathcal{S}$-matrix of the theory. If we wanted to obtain something similar from (\ref{ch4-eq1})--(\ref{ch4-eq2}) the usual strategy would be the expansion of the $\mathcal{S}$-matrix in the functional Taylor series with respect to the field configurations $\varphi$, and that is the meaning of the expression (\ref{ch2-eq19}). This expansion would generate an infinite hierarchy (chain) of coupled integro-differential equations for the corresponding functions $\mathcal{S}^{(n)}$, which are the expansion coefficients for the functional Taylor series. For example, the equation for a two-particle function $\mathcal{S}^{(2)}$ contains also a four-particle function $\mathcal{S}^{(4)}$ in special kinematics (for simplicity we assume that the functions of odd order $\mathcal{S}^{(n)}$ identically equal to zero which is correct for theories with an even in the field $\varphi$ interaction Lagrangian). If we continued the derivation of the equation for a four-particle function $\mathcal{S}^{(4)}$ it would contain a six-particle function $\mathcal{S}^{(6)}$ (in special kinematics) and so on. We can call it ``$n,\,n+2$ problem'', and this problem is the main difficulty in the FRG method.

\subsection{Different coarse-graining procedures of degrees of freedom}

In this subsection we pay our attention to the ambiguity of the FRG flow procedures (the original treatment can be found in \cite{rosten2012fundamentals}). For definiteness the Wilson--Polchinski equation (\ref{ch4-eq3}) or (\ref{ch4-eq4}) for the generating functional $\mathcal{G}$ is considered. This equation is a special case of a more general functional flow construction which can be formulated in a form of implicit (with respect to the functional $\mathcal{G}$) functional equation:
    \begin{eqnarray}\label{ch4-eq5}
        \frac{\delta\mathcal{G}\left[\varLambda,g,\varphi\right]}
        {\delta\varLambda}=\mathrm{Tr}
        \left\{\frac{\delta\varPsi_{\mathcal{G}}
        \left[\varLambda,g,\varphi\right]}
        {\delta\varphi}\right\}+
        \left(\frac{\delta\mathcal{G}
        \left[\varLambda,g,\varphi\right]}
        {\delta\varphi}\bigg|\varPsi_{\mathcal{G}}
        \left[\varLambda,g,\varphi\right]\right).
    \end{eqnarray}
In the coordinate representation the construction (\ref{ch4-eq5}) is given by:
    \begin{eqnarray}\label{ch4-eq6}
        \frac{\delta\mathcal{G}\left[\varLambda,g,\varphi\right]}
        {\delta\varLambda\left(z\right)}=
        \int d^{D}x\left\{\frac{\delta\varPsi_{\mathcal{G}}
        \left[\varLambda,g,\varphi\right]\left(z,x\right)}
        {\delta\varphi\left(x\right)}+
        \frac{\delta\mathcal{G}\left[\varLambda,g,\varphi\right]}
        {\delta\varphi\left(x\right)}\,\varPsi_{\mathcal{G}}
        \left[\varLambda,g,\varphi\right]\left(z,x\right)\right\}.
    \end{eqnarray}
The object $\varPsi$ has one ``index'' $x$ and is a functional of a field variable $\varphi$. Moreover, it depends on the functional $\mathcal{G}$ which makes the functional flow equations (\ref{ch4-eq5})--(\ref{ch4-eq6}) implicit. The meaning of $\varPsi$ is that this object parameterizes the process of increasing of coarse grain level for degrees of freedom in the system, in other words, one or another FRG flow procedure. At the same moment $\varPsi$ satisfies only general conditions and its specific form is up to a particular case. 

In order to get the Wilson--Polchinski equation (\ref{ch4-eq3})--(\ref{ch4-eq4}) from (\ref{ch4-eq5})--(\ref{ch4-eq6}) we need to make the following choice (in abstract form):
    \begin{eqnarray}\label{ch4-eq7}
        \varPsi_{\mathcal{G}}\left[\varLambda,g,\varphi\right]=\frac{1}{2}
        \frac{\delta\mathbf{G}\left[\varLambda\right]}{\delta\varLambda}\,
        \frac{\delta\mathcal{G}\left[\varLambda,g,\varphi\right]}
        {\delta\varphi}.
    \end{eqnarray}
In the coordinate representation the abstract expression (\ref{ch4-eq7}) reads:
    \begin{eqnarray}\label{ch4-eq8}
        \varPsi_{\mathcal{G}}\left[\varLambda,g,\varphi\right]\left(z,x\right)=\frac{1}{2}\int d^{D}y\,\frac{\delta G\left[\varLambda\right]\left(x,y\right)}
        {\delta\varLambda\left(z\right)}\frac{\delta\mathcal{G}
        \left[\varLambda,g,\varphi\right]}{\delta\varphi\left(y\right)}.
    \end{eqnarray}
Therefore, the equations (\ref{ch4-eq5})--(\ref{ch4-eq6}) demonstrate us a large functional ambiguity of the FRG method. Nevertheless, this ambiguity is not a drawback but an opportunity that may provide us a deeper insight into quantum field theory.

\subsection{Hamilton--Jacobi functional equation} 

In the final subsection we consider the functional Hamilton--Jacobi equation. This equation is the basic one in the holographic renormalization group method just like the functional Wilson--Polchinski equation is the basis for the one of the realizations of FRG method. It is worth noting that the Wilson--Polchinski equation is exact while the Hamilton--Jacobi equation represents some simplified model.

The standard approach widely used in literature (see, for instance, \cite{heemskerk2011holographic}) is that the Hamilton--Jacobi equation is formulated on the boundary of some extended with respect to $x$ spacetime (in our paper we use term ``space'' for brevity). We have already met an example of such a space when we saw the unification of $x$ and $\lambda$ in the expressions (\ref{ch2-eq20})--(\ref{ch2-eq21}) for the $\mathcal{S}$-matrix. The most common example of such a space for holographic renormalization group is anti-de Sitter one (papers \cite{maldacena1999large,witten1998anti,gubser1998gauge}).

In this paper we propose another model: let us formulate the theory in the $D$-dimensional $x$ space. Instead of additional coordinates (further we consider one extra dimension for simplicity) we consider a holographic field $\varLambda_{h}$ in the $D$-dimensional $x$ space. We can obtain an additional coordinate as a value of the delta-field (in terms of Dirac delta-function) configuration of the field $\varLambda_{h}$ (partially the idea comes from \cite{doplicher2004generalized}). The argumentation of such a formulation of the theory is the following: this approach allows us to formulate not only the Hamilton--Jacobi functional equation but also the whole hierarchy of the integro-differential equations for a (holographic) family of Green functions which generates by the functional which satisfies the functional Hamilton--Jacobi equation.

Let us illustrate this. Let the {\bf\emph{Hamilton functional}} be of the form ($z$ is the value of the $D$-dimensional coordinate, $\pi$ is the momentum of the field $\varphi$): 
    \begin{eqnarray}\label{ch4-eq9}
        \mathcal{H}\left[\varLambda_{h},\pi,\varphi\right]\left(z\right)=
        \mathcal{K}\left[\varLambda_{h},\pi\right]\left(z\right)+
        \mathcal{W}\left[\varLambda_{h},\varphi\right]\left(z\right).
    \end{eqnarray}
Imitating classical physics we chose the {\bf\emph{kinetic energy functional}} $\mathcal{K}$ in its simplest form:
    \begin{eqnarray}\label{ch4-eq10}
        \mathcal{K}\left[\varLambda_{h},\pi\right]\left(z\right)=
        \frac{1}{2}\int d^{D}x_{1}\int d^{D}x_{2}\,
        \mathcal{K}\left[\varLambda_{h}\right]
        \left(z,x_{1},x_{2}\right)\pi\left(x_{1}\right)
        \pi\left(x_{2}\right).
    \end{eqnarray}
The expression for the {\bf\emph{potential energy functional}} $\mathcal{W}$ is given by:
    \begin{eqnarray}\label{ch4-eq11}
        \mathcal{W}\left[\varLambda_{h},\varphi\right]\left(z\right)=
        \sum\limits_{n=2}^{\infty}\frac{1}{n!}\int d^{D}x_{1}\ldots\int d^{D}x_{n}\times\nonumber\\
        \times\mathcal{H}^{(n)}\left[\varLambda_{h}\right]
        \left(z,x_{1},\ldots,x_{n}\right)
        \varphi\left(x_{1}\right)\ldots\varphi\left(x_{n}\right).
    \end{eqnarray}
Using the Hamilton functional (\ref{ch4-eq9})--(\ref{ch4-eq11}) the corresponding Hamilton--Jacobi equation can be written in a standard way:
    \begin{eqnarray}\label{ch4-eq12}
        \frac{\delta\mathcal{G}\left[\varLambda_{h},\varphi\right]}
        {\delta\varLambda_{h}\left(z\right)}=\mathcal{H}\left[\varLambda_{h},\pi=\frac{\delta\mathcal{G}}{\delta\varphi},\varphi\right]\left(z\right).
    \end{eqnarray}
In the explicit form this equation reads:
    \begin{eqnarray}\label{ch4-eq13}
        \frac{\delta\mathcal{G}\left[\varLambda_{h},\varphi\right]}
        {\delta\varLambda_{h}\left(z\right)}=\frac{1}{2}
        \int d^{D}x_{1}\int d^{D}x_{2}\,\,
        \mathcal{K}\left[\varLambda_{h}\right]
        \left(z,x_{1},x_{2}\right)\times\nonumber\\
        \times\frac{\delta\mathcal{G}\left[\varLambda_{h},\varphi\right]}{\delta\varphi\left(x_{1}\right)}\frac{\delta\mathcal{G}\left[\varLambda_{h},\varphi\right]}{\delta\varphi\left(x_{2}\right)}+\mathcal{W}\left[\varLambda_{h},\varphi\right]\left(z\right).
    \end{eqnarray}
The main advantage of the equation (\ref{ch4-eq13}) in comparison with the FRG flow equations of Wilson--Polchinski and Wetterich--Morris is that this equation does not have ``$n,\,n+2$ problem''. Indeed, the equation for the $n$-particle (holographic) Green function will always contain Green functions of order $n'\leq n$.

As an example let us consider the equation for the two-particle Green function:
    \begin{eqnarray}\label{ch4-eq14}
        \frac{\delta\mathcal{G}^{\left(2\right)}
        \left[\varLambda_{h}\right]\left(y_{1},y_{2}\right)}
        {\delta\varLambda_{h}\left(z\right)}=\!\!\int\!\!d^{D}x_{1}
        \!\!\int\!\!d^{D}x_{2}\,\,\mathcal{K}\left[\varLambda_{h}\right]
        \left(z,x_{1},x_{2}\right)\times\nonumber\\
        \times\mathcal{G}^{\left(2\right)}
        \left[\varLambda_{h}\right]\left(y_{1},x_{1}\right)
        \mathcal{G}^{\left(2\right)}\left[\varLambda_{h}\right]
        \left(y_{2},x_{2}\right)+
        \mathcal{H}^{(2)}\left[\varLambda_{h}\right]
        \left(z,y_{1},y_{2}\right).\label{hologr6}
    \end{eqnarray}
Assuming in the expression (\ref{ch4-eq14}) the delta-field configurations $\varLambda_{h}\left(z\right)=\varLambda_{h}\delta_{z,0}$ (the value of the $D$-dimensional field argument can be chosen arbitrarily, in the general case, different from zero but zero value is enough for our consideration) we obtain a well-defined equation for the holographic version of the two-particle Green function of some quantum field theory. 

Let us recall that the Wilson--Polchinski equation (\ref{ch4-eq3})--(\ref{ch4-eq4}) is exact while the functional Hamilton--Jacobi equation (\ref{ch4-eq13}) represents some simplified model. Despite this, we can use the following observation. The right-hand side of the Wilson--Polchinski equation contains the sum of two terms. The term containing the square of the first derivative can be called ``classical'', since it depends only on the square of the functional momentum $\pi$. The functional momentum itself, as in the case of classical (analytical) mechanics, is defined as the first derivative of the functional $\mathcal{G}$ with respect to the ``generalized coordinate'' -- the field $\varphi$. The term containing the second derivative can be called ``quantum'' because it depends on the derivative of the functional momentum $\pi$ with respect to the field $\varphi$. Further, in the semiclassical approximation, the interaction action $S_{1}$ of the system with a minus sign can be substituted in the second term instead of the exact functional $\mathcal{G}$. In this case, we arrive at the Hamilton--Jacobi equation, and we can identify the holographic field $\varLambda_{h}$ with the field $\varLambda$. Let us note that \emph{the solution to the Hamilton--Jacobi equation, which is obtained in this way, should be used as the first step for the iterative procedure for solving the Wilson--Polchinski equation.} Moreover, having the exact solution for the two-particle Green function from the expressions (\ref{ch2-eq20})--(\ref{ch2-eq21}), obtained, for example, numerically, one can compare this solution with the solution for the two-particle Green function obtained by solving the Hamilton--Jacobi equation. Unfortunately, such a comparison is not possible analytically. Finally, we note that the holographic field $\varLambda_{h}$ can be chosen quite arbitrarily. If we represent it as a certain amplitude multiplied by the shape function $\varLambda_{h}\left(z\right)=\varLambda_{h}\,
\mathrm{shape}\left(z\right)$, it is the field amplitude $\varLambda_{h}$ that is the renormalization group scale (or additional coordinate).

\section{Conclusion}\label{ch5}

In this paper, we consider nonlocal and nonpolynomial QFT of the one component scalar field in $D$ dimensional spacetime, which was formulated by G.V. Efimov. Simultaneously, in the subsection entitled ``Zoology of the nonpolynomial Lagrangians in nonlocal QFT and nested exponentials'', we present the classification of different theories which can be examined by methods discussed in this paper. Also, we discuss the important duality of the structure of propagator in nonlocal QFT and the structure of nonlocal interaction Lagrangian. Meanwhile, in authors opinion, the central idea of this theory is the representation of interaction action $S_{1}$ in a form of {\bf\emph{exp in the power of exp}} under the path integral sign (for generating functional $\mathcal{Z}$ of total Green functions or for $\mathcal{S}$-matrix). Expanding the {\bf\emph{external exp}} (with argument $S_{1}$) into Taylor series, we obtain infinite series of Gaussian path integrals and each path integral can be calculated exactly. All the calculations are carried out in {\bf\emph{Euclidean}} metrics.

The expression for the path integral, which represent $\mathcal{S}$-matrix, is obtained in terms of the grand canonical partition function of $(D+N)$-dimensional {\bf\emph{classical}} interacting gas. In the present paper, $N=1$, in other words, there is one additional coordinate (extra dimension) $\lambda$ which emerged from the Fourier transform for interaction Lagrangian $U$. The physical meaning of the additional coordinate $\lambda$ (in general, set of additional coordinates $\lambda_1,\ldots,\lambda_N$) can be understood from the following considerations. In quantum field theory and statistical physics the field $\varphi$ plays the role of the argument of the functional. Therefore, $\varphi$ defines the domain of the theory. Space coordinate $x$ (the argument of $\varphi$) plays the role of the index. However, if the interaction Lagrangian $U$ has the specific form (\ref{ch2-eq5}) the domain of the theory is defined as the extension of space $x$ by the inclusion of additional coordinate $\lambda$. The value $\lambda$ is the spectrum of the field $\varphi$. Thus, in the theory under consideration, $\varphi$ has a meaning of an extra dimension more then a function of $x$. 

This duality can be very useful lesson for researchers in the field of AdS/CFT correspondence. However, the established duality does not answer the next question: How to obtain the grand canonical partition function in an explicit form? In the present paper, the toy model for realistic QFT is examined. In framework of this toy model, the $\mathcal{S}$-matrix can be calculated in a closed form but it is only a toy. For the solution of realistic QFT problem it may be useful to analyze different functional equations (in general, with functional derivatives) in Efimov representation. In this representation, Schwinger--Dyson and functional Schr\"{o}dinger equations for different generating functionals are derived in details (in completely integral form). Further in the present paper several asymptotic solutions of Schwinger--Dyson equation for $\mathcal{S}$-matrix are derived. Also, the solutions of self-consistency equations are obtained. These equations demonstrate the non-triviality of the constructed $\mathcal{S}$-matrix. A part of the final results is illustrated in context of $D$-dimensional sine-Gordon theory but obtained conclusions usually are true in general.

In the story about functional equations, ultraviolet form factors and holography several famous equations of QFT are briefly discussed. They are the Wilson--Polchinski equation, the Wetterich--Morris equation, including discussion of the different coarse-graining procedures of degrees of freedom (these procedures define the explicit form of FRG flow equations), and the equation of the holographic renormalization group, i.e. the functional Hamilton--Jacobi equation. A distinctive feature of the applied approach for holographic renormalization group is that we formulate the Hamilton--Jacobi equation in $D$-dimensional real (coordinate) space $x$ but in the presence of the additional {\bf\emph{holographic}} scalar field. The delta-field configurations of this field create an additional (holographic) coordinate. This approach allows us to formulate not only the Hamilton--Jacobi equation but all the coupled hierarchy of integro-differential equations for the set of holographic Green functions. We also note that the obtained hierarchy is {\bf\emph{decomposed}}.

In conclusion, we should note further possible research directions. First of all, one can try to sum the expression for the $\mathcal{S}$-matrix in terms of the grand canonical partition function of $(D+N)$-dimensional interacting classical gas. There is an advanced mathematical technique of such fundamental sciences as statistical physics and physical kinetics for this purpose. Alternatively, it is possible to try to construct the solutions to the Schwinger--Dyson equation (for instance) for the $\mathcal{S}$-matrix in terms of the ansatz, which includes obtained asymptotic expressions for the $\mathcal{S}$-matrix in the limits of large and small fields. In other words, it is possible to obtain solutions using the perturbation theory around the asymptotics obtained in this paper. In authors view this research direction is the most interesting because these series will be an alternative expansion (for example) for $\mathcal{S}$-matrix in contrast to expressions obtained by Efimov. Finally, one can consider different field configurations (arguments of the $\mathcal{S}$-matrix), include different composite operators, find different families of Green functions of the theory and analyze the dependence of the obtained expressions on the ultraviolet parameter $l$. Here we note that the Euclidean metric is used in all mentioned problems. Therefore, there is an interesting and difficult problem of analytic continuation of the obtained results to Minkowski space. Fortunately, questions such as the unitarity of the obtained $\mathcal{S}$-matrix have positive answers due to G.V. Efimov's papers. Thus, methods of nonlocal and nonpolynomial QFT are able to take a worthy place in the description of Nature.

\section*{Acknowledgments}

The authors are deeply grateful to their families for love, wisdom and understanding. We are very grateful to Artem A. Alexandrov for his help in typing the paper. Also, we express special gratitude to Sergey E. Kuratov and Alexander V. Andriyash for supporting this research at an early stage at the Center for Fundamental and Applied Research (Dukhov Research Institute of Automatics). Finally, we are very grateful to Reviewer for many valuable comments and advice on this paper.

\end{document}